%% file: ng12p5yr_custom_noise.tex
\documentclass[twocolumn]{aastex7}

\submitjournal{ApJ}

\shorttitle{NANOGrav Time-Domain Chromatic Noise}
\shortauthors{The NANOGrav Collaboration}

\usepackage{xspace}
\usepackage{rotating,booktabs}
\usepackage{multirow}
\usepackage{amsmath}

\graphicspath{{./}{./figures/}}

\newcommand{\mysec}[1]{\S\ref{#1}\xspace}
\newcommand{\myfig}[1]{Figure~\ref{#1}\xspace}
\newcommand{\mytab}[1]{Table~\ref{#1}\xspace}
\newcommand{\myeq}[1]{Equation~\ref{#1}\xspace}

\newcommand{\psrplus}[2]{PSR #1+#2\xspace}
\newcommand{\psrminus}[2]{PSR #1$-$#2\xspace}
\newcommand{\pplus}[2]{#1+#2\xspace}
\newcommand{\pminus}[2]{#1$-$#2\xspace}

\defcitealias{NG11a}{NG11a}
\defcitealias{NG12p5data}{NG12p5data}
\defcitealias{NG12p5gwb}{NG12p5gwb}

\begin{document}

\title{The NANOGrav $12.5$-year Data Set:\\Chromatic Noise Characterization \& Mitigation with Time-Domain Kernels}

\author[0000-0003-2742-3321]{Jeffrey S. Hazboun}
\affiliation{Department of Physics, Oregon State University, Corvallis, OR 97331, USA}
\email{jeffrey.hazboun@oregonstate.edu}
\correspondingauthor{Jeffrey S. Hazboun}
\author[0000-0003-1407-6607]{Joseph Simon} 
\affiliation{Department of Physics, Oregon State University, Corvallis, OR 97331, USA}
\affiliation{Department of Astrophysical and Planetary Sciences, University of Colorado, Boulder, CO 80309, USA}
\email{joseph.simon@nanograv.org}
\author[0000-0002-4972-1525]{Jeremy Baier}
\affiliation{Department of Physics, Oregon State University, Corvallis, OR 97331, USA}
\email{jeremy.baier@nanograv.org}
\author[0000-0001-6436-8216]{Bjorn Larsen}
\affiliation{Department of Physics, Yale University, New Haven, CT 06520, USA}
\email{bjorn.larsen@nanograv.org}
\author[0000-0002-7374-6925]{Daniel J. Oliver}
\affiliation{Department of Physics, Oregon State University, Corvallis, OR 97331, USA}
\email{daniel.oliver@nanograv.org}
\author[0000-0003-2745-753X]{Paul T. Baker}
\affiliation{Department of Physics and Astronomy, Widener University, One University Place, Chester, PA 19013, USA}
\email{paul.baker@nanograv.org}
\author[0000-0003-0909-5563]{Bence B\'{e}csy}
\affiliation{Institute for Gravitational Wave Astronomy and School of Physics and Astronomy, University of Birmingham, Edgbaston, Birmingham B15 2TT, UK}
\email{bence.becsy@nanograv.org}
\author[0000-0002-3118-5963]{Siyuan Chen}
\affiliation{Shanghai Astronomical Observatory, Chinese Academy of Sciences, 80 Nandan Road, Shanghai 200030, China}
\affiliation{State Key Laboratory of Radio Astronomy and Technology, A20 Datun Road, Beijing 100101, China}
\email{siyuan.chen@nanograv.org}
\author[0009-0009-0304-6753]{Alberto Diaz Hernandez}
\affiliation{Department of Physics, Oregon State University, Corvallis, OR 97331, USA}
\email{diazhera@oregonstate.edu}
\author{Justin A. Ellis}
\affiliation{Infinia ML, 202 Rigsbee Avenue, Durham NC, 27701}
\email{justin.ellis@nanograv.org}
\author{A. Miguel Holgado}
\affiliation{Department of Astronomy and National Center for Supercomputing Applications, University of Illinois at Urbana-Champaign, Urbana, IL 61801, USA}
\email{miguel.holgado@nanograv.org}
\author{Kristina Islo}
\affiliation{Center for Gravitation, Cosmology and Astrophysics, Department of Physics, University of Wisconsin-Milwaukee,\\ P.O. Box 413, Milwaukee, WI 53201, USA}
\email{kristina.islo@nanograv.org}
\author[0000-0002-7445-8423]{Aaron Johnson}
\affiliation{Center for Gravitation, Cosmology and Astrophysics, Department of Physics, University of Wisconsin-Milwaukee,\\ P.O. Box 413, Milwaukee, WI 53201, USA}
\affiliation{Division of Physics, Mathematics, and Astronomy, California Institute of Technology, Pasadena, CA 91125, USA}
\email{aaron.johnson@nanograv.org}
\author[0000-0002-3654-980X]{Andrew R. Kaiser}
\affiliation{Department of Physics and Astronomy, West Virginia University, P.O. Box 6315, Morgantown, WV 26506, USA}
\email{andrew.kaiser@nanograv.org}
\author[0000-0002-9197-7604]{Nima Laal}
\affiliation{Department of Physics and Astronomy, Vanderbilt University, 2301 Vanderbilt Place, Nashville, TN 37235, USA}
\email{nima.laal@nanograv.org}
\author[0000-0001-5481-7559]{Alexander McEwen}
\affiliation{Center for Gravitation, Cosmology and Astrophysics, Department of Physics, University of Wisconsin-Milwaukee,\\ P.O. Box 413, Milwaukee, WI 53201, USA}
\email{alexander.mcewen@nanograv.org}
\author[0000-0002-8826-1285]{Nihan S. Pol}
\affiliation{Department of Physics, Texas Tech University, Box 41051, Lubbock, TX 79409, USA}
\email{nihan.pol@nanograv.org}
\author[0000-0003-0123-7600]{Joey Shapiro Key}
\affiliation{University of Washington Bothell, 18115 Campus Way NE, Bothell, WA 98011, USA}
\email{joey.key@nanograv.org}
\author{Min Young Kim}
\affiliation{University of Washington Bothell, 18115 Campus Way NE, Bothell, WA 98011, USA}
\email{min.young.kim@nanograv.org}
\author{Matthew Samson}
\affiliation{Department of Physics and Astronomy, Widener University, One University Place, Chester, PA 19013, USA}
\email{matthew.samson@nanograv.org}
\author[0000-0002-7283-1124]{Brent J. Shapiro-Albert}
\affiliation{Department of Physics and Astronomy, West Virginia University, P.O. Box 6315, Morgantown, WV 26506, USA}
\affiliation{Center for Gravitational Waves and Cosmology, West Virginia University, Chestnut Ridge Research Building, Morgantown, WV 26505, USA}
\affiliation{Giant Army, 915A 17th Ave, Seattle WA 98122}
\email{brent.shapiro-albert@nanograv.org}
\author[0000-0002-7933-493X]{Jerry P. Sun}
\affiliation{Department of Physics, Oregon State University, Corvallis, OR 97331, USA}
\email{jerry.sun@nanograv.org}
\author[0000-0003-0264-1453]{Stephen R. Taylor}
\affiliation{Department of Physics and Astronomy, Vanderbilt University, 2301 Vanderbilt Place, Nashville, TN 37235, USA}
\email{stephen.taylor@nanograv.org}
\author[0000-0002-6020-9274]{Caitlin A. Witt}
\affiliation{Center for Interdisciplinary Exploration and Research in Astrophysics (CIERA), Northwestern University, Evanston, IL 60208, USA}
\affiliation{Adler Planetarium, 1300 S. DuSable Lake Shore Dr., Chicago, IL 60605, USA}
\email{caitlin.witt@nanograv.org}
\author{Jeremy Volpe}
\affiliation{Department of Physics and Astronomy, Widener University, One University Place, Chester, PA 19013, USA}
\email{jeremy.volpe@nanograv.org}
\author[0000-0002-8559-0788]{Christine Ye}
\affiliation{Department of Mathematics, Stanford University, USA}
\email{cye@stanford.edu}


\author[0000-0003-4046-884X]{Harsha Blumer}
\affiliation{Department of Physics and Astronomy, West Virginia University, P.O. Box 6315, Morgantown, WV 26506, USA}
\affiliation{Center for Gravitational Waves and Cosmology, West Virginia University, Chestnut Ridge Research Building, Morgantown, WV 26505, USA}
\email{harsha.blumer@nanograv.org}
\author[0000-0003-3053-6538]{Paul R. Brook}
\affiliation{Institute for Gravitational Wave Astronomy and School of Physics and Astronomy, University of Birmingham, Edgbaston, Birmingham B15 2TT, UK}
\affiliation{Center for Gravitational Waves and Cosmology, West Virginia University, Chestnut Ridge Research Building, Morgantown, WV 26505, USA}
\email{paul.brook@nanograv.org}
\author[0000-0002-2878-1502]{Shami Chatterjee}
\affiliation{Cornell Center for Astrophysics and Planetary Science and Department of Astronomy, Cornell University, Ithaca, NY 14853, USA}
\email{shami.chatterjee@nanograv.org}
\author[0000-0002-4049-1882]{James M. Cordes}
\affiliation{Cornell Center for Astrophysics and Planetary Science and Department of Astronomy, Cornell University, Ithaca, NY 14853, USA}
\email{james.cordes@nanograv.org}
\author[0000-0002-2578-0360]{Fronefield Crawford}
\affiliation{Department of Physics and Astronomy, Franklin \& Marshall College, P.O. Box 3003, Lancaster, PA 17604, USA}
\email{fronefield.crawford@nanograv.org}
\author[0000-0002-6039-692X]{H. Thankful Cromartie}
\affiliation{National Research Council Research Associate, National Academy of Sciences, Washington, DC 20001, USA resident at Naval Research Laboratory, Washington, DC 20375, USA}
\email{thankful.cromartie@nanograv.org}
\author[0000-0002-2185-1790]{Megan E. DeCesar}
\affiliation{Department of Physics and Astronomy, George Mason University, Fairfax, VA 22030, resident at the U.S. Naval Research Laboratory, Washington, DC 20375, USA}
\altaffiliation{Resident at the Naval Research Laboratory}
\email{megan.decesar@nanograv.org}
\author[0000-0002-6664-965X]{Paul B. Demorest}
\affiliation{National Radio Astronomy Observatory, 1003 Lopezville Rd., Socorro, NM 87801, USA}
\email{paul.demorest@nanograv.org}
\author[0000-0001-8885-6388]{Timothy Dolch}
\affiliation{Department of Physics, Hillsdale College, 33 E. College Street, Hillsdale, Michigan 49242, USA}
\affiliation{Eureka Scientific, 2452 Delmer Street, Suite 100, Oakland, CA 94602-3017, USA}
\affiliation{SETI Institute, 339 N Bernardo Ave Suite 200, Mountain View, CA 94043, USA}
\email{timothy.dolch@nanograv.org}
\author[0000-0002-2223-1235]{Robert D. Ferdman}
\affiliation{School of Chemistry, University of East Anglia, Norwich, NR4 7TJ, United Kingdom}
\email{rob.ferdman@nanograv.org}
\author{Elizabeth C. Ferrara}
\affiliation{NASA Goddard Space Flight Center, Greenbelt, MD 20771, USA}
\affiliation{Department of Astronomy, University of Maryland, College Park, MD 20742, USA}
\email{elizabeth.ferrara@nanograv.org}
\author[0000-0001-5645-5336]{William Fiore}
\affiliation{Department of Physics and Astronomy, University of British Columbia, 6224 Agricultural Road, Vancouver, BC V6T 1Z1, Canada}
\email{william.fiore@nanograv.org}
\author[0000-0001-8384-5049]{Emmanuel Fonseca}
\affiliation{Department of Physics and Astronomy, West Virginia University, P.O. Box 6315, Morgantown, WV 26506, USA}
\affiliation{Center for Gravitational Waves and Cosmology, West Virginia University, Chestnut Ridge Research Building, Morgantown, WV 26505, USA}
\email{emmanuel.fonseca@nanograv.org}
\author[0000-0001-6166-9646]{Nathan Garver-Daniels}
\affiliation{Department of Physics and Astronomy, West Virginia University, P.O. Box 6315, Morgantown, WV 26506, USA}
\affiliation{Center for Gravitational Waves and Cosmology, West Virginia University, Chestnut Ridge Research Building, Morgantown, WV 26505, USA}
\email{nate.garver-daniels@nanograv.org}
\author[0000-0001-8158-683X]{Peter A. Gentile}
\affiliation{Department of Physics and Astronomy, West Virginia University, P.O. Box 6315, Morgantown, WV 26506, USA}
\affiliation{Center for Gravitational Waves and Cosmology, West Virginia University, Chestnut Ridge Research Building, Morgantown, WV 26505, USA}
\email{peter.gentile@nanograv.org}
\author[0000-0003-1884-348X]{Deborah C. Good}
\affiliation{Department of Physics and Astronomy, University of Montana, 32 Campus Drive, Missoula, MT 59812}
\email{deborah.good@nanograv.org}
\author[0000-0003-1082-2342]{Ross J. Jennings}
\affiliation{Department of Physics and Astronomy, West Virginia University, P.O. Box 6315, Morgantown, WV 26506, USA}
\affiliation{Center for Gravitational Waves and Cosmology, West Virginia University, Chestnut Ridge Research Building, Morgantown, WV 26505, USA}
\altaffiliation{NANOGrav Physics Frontiers Center Postdoctoral Fellow}
\email{ross.jennings@nanograv.org}
\author[0000-0001-6607-3710]{Megan L. Jones}
\affiliation{Center for Gravitation, Cosmology and Astrophysics, Department of Physics, University of Wisconsin-Milwaukee,\\ P.O. Box 413, Milwaukee, WI 53201, USA}
\affiliation{Center for Gravitational Waves and Cosmology, West Virginia University, Chestnut Ridge Research Building, Morgantown, WV 26505, USA}
\email{megan.jones@nanograv.org}
\author[0000-0001-6295-2881]{David L. Kaplan}
\affiliation{Center for Gravitation, Cosmology and Astrophysics, Department of Physics, University of Wisconsin-Milwaukee,\\ P.O. Box 413, Milwaukee, WI 53201, USA}
\email{david.kaplan@nanograv.org}
\author[0000-0003-0721-651X]{Michael T. Lam}
\affiliation{SETI Institute, 339 N Bernardo Ave Suite 200, Mountain View, CA 94043, USA}
\affiliation{Laboratory for Multiwavelength Astrophysics, Rochester Institute of Technology, Rochester, NY 14623, USA}
\affiliation{School of Physics and Astronomy, Rochester Institute of Technology, Rochester, NY 14623, USA}
\email{michael.lam@nanograv.org}
\author{T. Joseph W. Lazio}
\affiliation{Jet Propulsion Laboratory, California Institute of Technology, 4800 Oak Grove Drive, Pasadena, CA 91109, USA}
\email{joseph.lazio@nanograv.org}
\author[0000-0003-1301-966X]{Duncan R. Lorimer}
\affiliation{Department of Physics and Astronomy, West Virginia University, P.O. Box 6315, Morgantown, WV 26506, USA}
\affiliation{Center for Gravitational Waves and Cosmology, West Virginia University, Chestnut Ridge Research Building, Morgantown, WV 26505, USA}
\email{duncan.lorimer@nanograv.org}
\author{Jing Luo}
\affiliation{Department of Astronomy \& Astrophysics, University of Toronto, 50 Saint George Street, Toronto, ON M5S 3H4, Canada}
\email{jing.luo@nanograv.org}
\author[0000-0001-5229-7430]{Ryan S. Lynch}
\affiliation{Green Bank Observatory, P.O. Box 2, Green Bank, WV 24944, USA}
\email{ryan.lynch@nanograv.org}
\author[0000-0003-2285-0404]{Dustin R. Madison}
\affiliation{Department of Physics, Occidental College, 1600 Campus Road, Los Angeles, CA 90041, USA}
\email{dustin.madison@nanograv.org}
\author[0000-0001-7697-7422]{Maura A. McLaughlin}
\affiliation{Department of Physics and Astronomy, West Virginia University, P.O. Box 6315, Morgantown, WV 26506, USA}
\affiliation{Center for Gravitational Waves and Cosmology, West Virginia University, Chestnut Ridge Research Building, Morgantown, WV 26505, USA}
\email{maura.mclaughlin@nanograv.org}
\author[0000-0002-4307-1322]{Chiara M. F. Mingarelli}
\affiliation{Department of Physics, Yale University, New Haven, CT 06520, USA}
\email{chiara.mingarelli@nanograv.org}
\author[0000-0002-3616-5160]{Cherry Ng}
\affiliation{Dunlap Institute for Astronomy and Astrophysics, University of Toronto, 50 St. George St., Toronto, ON M5S 3H4, Canada}
\email{cherry.ng@nanograv.org}
\author[0000-0002-6709-2566]{David J. Nice}
\affiliation{Department of Physics, Lafayette College, Easton, PA 18042, USA}
\email{david.nice@nanograv.org}
\author[0000-0001-5465-2889]{Timothy T. Pennucci}
\affiliation{Institute of Physics, E\"{o}tv\"{o}s Lor\'{a}nd University, P\'{a}zm\'{a}ny P. s. 1/A, 1117 Budapest, Hungary}
\email{timothy.pennucci@nanograv.org}
\author[0000-0001-5799-9714]{Scott M. Ransom}
\affiliation{National Radio Astronomy Observatory, 520 Edgemont Road, Charlottesville, VA 22903, USA}
\email{scott.ransom@nanograv.org}
\author[0000-0002-5297-5278]{Paul S. Ray}
\affiliation{Space Science Division, Naval Research Laboratory, Washington, DC 20375-5352, USA}
\email{paul.ray@nanograv.org}
\author[0000-0002-7778-2990]{Xavier Siemens}
\affiliation{Department of Physics, Oregon State University, Corvallis, OR 97331, USA}
\affiliation{Center for Gravitation, Cosmology and Astrophysics, Department of Physics, University of Wisconsin-Milwaukee,\\ P.O. Box 413, Milwaukee, WI 53201, USA}
\email{xavier.siemens@nanograv.org}
\author[0000-0002-6730-3298]{Ren\'{e}e Spiewak}
\affiliation{Jodrell Bank Centre for Astrophysics, University of Manchester, Manchester, M13 9PL, United Kingdom}
\email{renee.spiewak@nanograv.org}
\author[0000-0001-9784-8670]{Ingrid H. Stairs}
\affiliation{Department of Physics and Astronomy, University of British Columbia, 6224 Agricultural Road, Vancouver, BC V6T 1Z1, Canada}
\email{ingrid.stairs@nanograv.org}
\author[0000-0002-1797-3277]{Daniel R. Stinebring}
\affiliation{Department of Physics and Astronomy, Oberlin College, Oberlin, OH 44074, USA}
\email{daniel.stinebring@nanograv.org}
\author[0000-0002-7261-594X]{Kevin Stovall}
\affiliation{National Radio Astronomy Observatory, 1003 Lopezville Rd., Socorro, NM 87801, USA}
\email{kevin.stovall@nanograv.org}
\author[0000-0002-1075-3837]{Joseph K. Swiggum}
\altaffiliation{NANOGrav Physics Frontiers Center Postdoctoral Fellow}
\affiliation{Department of Physics, Lafayette College, Easton, PA 18042, USA}
\email{joseph.swiggum@nanograv.org}
\author[0000-0002-2451-7288]{Jacob E. Turner}
\affiliation{Green Bank Observatory, P.O. Box 2, Green Bank, WV 24944, USA}
\email{jacob.turner@nanograv.org}
\author[0000-0002-4162-0033]{Michele Vallisneri}
\affiliation{Jet Propulsion Laboratory, California Institute of Technology, 4800 Oak Grove Drive, Pasadena, CA 91109, USA}
\email{michele.vallisneri@nanograv.org}
\author[0000-0003-4700-9072]{Sarah J. Vigeland}
\affiliation{Center for Gravitation, Cosmology and Astrophysics, Department of Physics, University of Wisconsin-Milwaukee,\\ P.O. Box 413, Milwaukee, WI 53201, USA}
\email{sarah.vigeland@nanograv.org}

\begin{abstract}
Pulsar timing arrays (PTAs) have recently entered the detection era, quickly moving beyond the goal of simply improving sensitivity at the lowest frequencies for the sake of observing the stochastic gravitational wave background (GWB), and focusing on its accurate spectral characterization. While all PTA collaborations around the world use Fourier-domain Gaussian processes to model the GWB and intrinsic long time-correlated (red) noise, techniques to model the time-correlated radio frequency-dependent (chromatic) processes have varied from collaboration to collaboration. Here we test a new class of models for PTA data, Gaussian processes based on time-domain kernels that model the statistics of the chromatic processes starting from the covariance matrix. As we will show, these models can be effectively equivalent to Fourier-domain models in mitigating chromatic noise. This work presents a method for Bayesian model selection across the various choices of kernel as well as deterministic chromatic models for non-stationary chromatic events and the solar wind. As PTAs turn towards high frequency ($>1/{\rm yr}$) sensitivity, the size of the basis used to model these processes will need to increase, and these time-domain models present some computational efficiencies compared to Fourier-domain models. 
\end{abstract}


\section{Introduction} 
\label{sec:intro}

Millisecond pulsars (MSPs) are stable astrophysical clocks uniquely suited to be used in fundamental experiments \citep{taylor:binarypulsars1993}, in particular, the detection of low-frequency (nHz-$\mu$Hz) gravitational waves (GWs) \citep{foster+1990}. Recently, pulsar timing array (PTA) experiments have uncovered evidence for a stochastic gravitational wave background (GWB) \citep{eptadr2_3:gwb,pptadr3:gwb,ng15gwb, ipta3p+2024, cptadr1_1:gwb, mpta4.5:gwb}. 

The GWB appears in PTA data as a common red noise (RN) signal inducing deviations from each pulsar's individual timing model, which accounts for the changing pulsar-observatory line of sight. In addition, there are radio-frequency dependent deviations induced by interstellar propagation effects. The GWB also induces a specific, quasi-quadrupolar angular correlation signature between the pulsars of an array, predicted by general relativity \citep{hd83}. 

Although the standard noise model used in NANOGrav analyses has been shown to be more than sufficient to isolate the cross-correlation pattern indicative of a GWB \citep{ng15detchar}, it has also been known for some time that there remains excess noise in individual pulsar data sets. Much of this noise appears chromatic in nature \citep{lcc+2017}, contaminating the timing model and reducing each pulsar's contribution to the detection of a GW signal. 
Additionally, GW signals are highly covariant with intrinsic pulsar red noise \citep{hazboun:2020slice}, which can be produced by a number of factors, including spin noise \citep{shannon_cordes_2010} and dispersion measure (DM) variations caused by the interstellar medium \citep{jones+2017_ng9_dm}. In the case of DM variations, we have additional information -- the dependence on radio frequency -- which makes it possible to disentangle these effects from GWs. Yet, at the same time, imperfect estimation of DM can be its own source of additional red noise \citep{shannon_cordes_2010}. 

In this paper, we introduce time-domain covariance functions for the stochastic time variations of chromatic time delays within PTA data. At the same time we investigate how the replacement of the current chromatic variations model, see DMX \citep{jones+2017_ng9_dm}, has on the data set.

Note that throughout this paper, when we refer to frequency, $f$, we are talking about the fluctuation frequency of the TOAs, i.e., the frequencies that would be associated with a Fourier transform of the timing residuals and what would be associated with GW frequencies. We will explicitly refer to ``radio frequency'', $\nu$, when we discuss the observation frequency of the TOAs at a radio telescope.

\section{Data}

NANOGrav's $12.5$-year data set includes observations of 47 pulsars made between July 2004 and June 2017 \citep{arz:2020}. All observations were taken with either the 305-m Arecibo Observatory (AO) or the 100-m Green Bank Telescope (GBT). AO was utilized for all pulsars that lie within its declination range ($0^\circ < \delta < +~39^\circ$), while GBT observed the pulsars outside of that declination range, in addition to \psrplus{J1713}{0747} and \psrplus{B1937}{21}. Most pulsars were observed approximately once per month, with six pulsars observed weekly as part of a high-cadence campaign that started in 2013 at GBT and 2015 at AO. When possible, pulsars were observed with two different wide-band receivers covering one higher and one lower radio frequency range in order to have sufficient sensitivity to pulse dispersion due to the interstellar medium (ISM). At AO, pulsars were observed using the 1.4 GHz receiver plus either the 430 MHz or 2.1 GHz receiver, depending on the pulsar's timing characteristics.\footnote{Early observations of \psrplus{J2317}{1439} also used the 327 MHz receiver.} At GBT, observations were taken using the 1.4 GHz and 800 MHz receivers. However, the separate frequency ranges were not observed simultaneously; instead, a few days separated the observations. A detailed discussion of the data set can be found in \citet{arz:2020}. 

The standard model used for time-varying chromatic delays in this data set, and all NANOGrav data sets, is referred to as DMX. The DMX model fits for variations to the DM, the integrated electron density along the line of sight to a pulsar, using all times of arrival within a set bin of time \citep{jones+2017_ng9_dm}. The bin sizes vary from $\sim 1$ hour to $1$ week long and are tuned to fit at least two receiver bands of times of arrival (TOAs) in a given bin to increase the precision of the time-varying DM.

In this work, we investigate the chromatic noise characterization for 20 of those pulsars (see \mytab{tab:modelresults}).
These pulsars were chosen due to a combination of a variety of factors: including their sensitivity to the common red noise process (described by their dropout factor, see \citet{arz+2020gwb} for more information), their observing timespan, and the presence of known excess noise features, among others. Ultimately, this work is not an exhaustive search of the entire $12.5$-year data set, but rather, it is an exploration of new time-domain chromatic models as well as an introduction to new methods for Bayesian model selection.

\section{Models}
\label{sec:models}

Gaussian processes (GPs) have been used in PTA data analysis for years \citep{lentati+:2012,vanhaasteren+2014_gp}, and are now a common feature of PTA data analysis pipelines \citep{ng15pipeline}. In this context, GPs are used to model pulsar timing residuals, stochastic noise processes in the data, and the GWB. Historically, Fourier-domain GP kernels have been used, which we review in \mysec{sec:fd_models}. This work also introduces time-domain kernels in \mysec{sec:td_models}, which we apply to model chromatic processes such as DM variations. The commonality between both implementations is a rank-reduced formalism where the signal model for our timing residuals, $s(t)$, is represented as a linear model $\vec{s}(t) = \mathbf{F}\vec{a}$, where $\vec{a}$ is a set of coefficients of the model and $\mathbf{F}$ is a design matrix mapping said coefficients to the full time-domain signal via a set of basis functions. The full-rank time-domain covariance of the signals is then $\mathbf{C} = \mathbf{F}\phi\mathbf{F}^T$, where the $\phi$ matrix gives the reduced-rank covariance of the coefficients ($\phi = \langle\vec{a}\vec{a}^T\rangle$), leading to more efficient matrix inversions via the Woodbury matrix identity. Throughout this work, $\mathbf{F}$ and $\phi$ may be defined differently depending on the model.

\subsection{Fourier-domain models}
\label{sec:fd_models}

Timing residuals induced by an astrophysical GWB should have a power spectral density (PSD) approximately following a power-law form,
\begin{equation}
    S(f) = \frac{{A^2}}{12\pi^2f_{\mathrm{yr}}^3} \left(\frac{f}{f_{\mathrm{yr}}}\right)^{\!\!-\gamma}\, \frac{\mathrm{s}^2}{\mathrm{Hz}},
\label{eq:powerlaw}
\end{equation}
where $A$ and $\gamma$ are the spectral amplitude and index parameters, $f$ is spectral frequency, and $f_{\mathrm{yr}}$ is the inverse of 1 year. Any generic $\gamma > 1$ signal may be considered a red noise process, while $\gamma_{\mathrm{GWB}} = 13/3$ is the expected value for a GWB from a population of circular supermassive black hole binaries \citep{Phinney2001}. As such, it is standard in PTA data analyses to model red noise as a GP using a truncated Fourier basis with coefficients drawn from the power-law PSD as a prior function. Under the rank-reduced formalism, $\mathbf{F}$ is a \emph{Fourier} design matrix with basis elements
\begin{align}
    \label{eq:Fmat}
    F_{ij} &= \begin{cases}
        \cos(2\pi f_{j/2}t_i) & \text{for even } j, \\
        \sin(2\pi f_{(j-1)/2}t_i) & \text{for odd } j,
    \end{cases}
\end{align}
given a set of linearly spaced frequencies $f_j = (j+1)/T$ for $j \in [0, N_f-1]$, where $T$ is the timespan of the data set and $N_f/T$ is a high frequency cutoff. Meanwhile, $\vec{a}$ encodes a set of Fourier coefficients, and the PSD is used to set priors on their variances, $\phi_{kl} = S(f)\delta_{kl}/T$, allowing use of \myeq{eq:powerlaw} for GP modeling of signals and noise. Another common choice is the ``free spectral'' prior, where the variances at each distinct frequency are sampled as independent parameters \citep{hazboun:2020slice}. These models assume no correlations between coefficients or frequencies, i.e., they are diagonal, though very recent work \citep{crisostomi+2025} has begun to look at non-diagonal kernels in the Fourier domain.

For achromatic noise, we typically set $N_f = 30$ as this covers the expected frequency range for low-frequency GWs, and the power law most often decays below the white noise (WN) level beyond $N_f > 30$ \citep{ng15detchar}. Additionally, the power-law Fourier basis GPs have been used extensively for modeling chromatic noise such as DM variations, implemented by including a $\nu$-dependent term in the basis (e.g., \citealt{lentati+:2016, goncharov+2021gwb, Chalumeau+2022, Miles+2023, Larsen+2024, Iraci+2024}). However, chromatic signals can exhibit complicated trends over time, which may not be straightforward to model in the Fourier domain, meaning either deviations from a power law form or very large values of $N_f$ may be required for accurate modeling \citep{cptadr1:singlepulsarnoise}. To account for this and introduce additional flexibility for chromatic noise modeling, we next introduce GPs whose bases and priors are defined in the time domain.

\subsection{Time-domain models}
\label{sec:td_models}
Some processes have temporal correlations that can straightforwardly be modeled directly in the time domain. To do so, we define a coarse-grained time-domain basis with nodes that are separated by a fixed number of days ($dt$). Processes defined in this coarse-grained basis are transformed to the full-rank time series through linear interpolation. To illustrate, consider our signal $s(t_i)$ at the ToA $t_i$ as the interpolation of the underlying stochastic signal $\vec{a}$ between preceding time $t'_{j_1}$ and a following time $t'_{j_2}$ with spacing $dt$,
\begin{align}
    \label{eq:interpolation}
    s(t_i) = \frac{(t'_{j_2} - t_i)a_{j_1} + (t_i - t'_{j_1})a_{j_2}}{dt}.
\end{align}
Since $\vec{s}(t)$ is linear with respect to $\vec{a}$, it is given as the matrix equation $\vec{s}(t) = \mathbf{F}\vec{a}$ where $\mathbf{F}$ is now the \emph{linear interpolation} design matrix (as opposed to the Fourier design matrix). Defining our linearly spaced grid $t_j' = \min(t) + j\cdot dt$ for $j \in [0,\lceil T/dt \rceil]$, $\mathbf{F}$ follows from \myeq{eq:interpolation} as
\begin{align}
    \label{eq:interp_basis}
    F_{ij} = \frac{1}{dt}\begin{cases}
        t'_{j+1} - t_i &\text{if } t'_j \le t_i \le t'_{j+1}, \\
        t_i - t_{j-1}' &\text{if } t'_{j-1} \le t_i \le t'_j, \\
        0 &\text{elsewhere}.
    \end{cases}
\end{align}
Additionally, we remove any empty columns of $\mathbf{F}$ that result when gaps between TOAs are larger than $2dt$. Here, the hyperparameter $dt$ plays a similar role as $N_f$ in the Fourier basis by determining both the high-frequency resolution of the model as well as the number of basis functions and resulting computational cost. Our default basis is $dt = 15$ days, but some pulsars require more fine-grained bases, as discussed in \mysec{sec:basis_size}.

\begin{figure}
    \centering
    \includegraphics[width=0.45\textwidth]{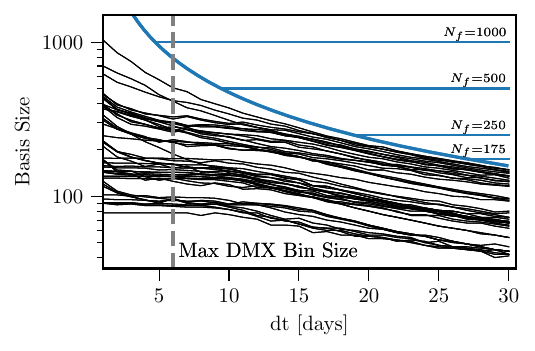}
    \caption{Time-domain basis size compared to Fourier-domain basis size. The blue line indicates the scaling of the Fourier basis size vs $dt$, given the high-frequency cutoff $f_{\mathrm{max}} = 2/dt$ and $N_f = f_{\mathrm{max}}T$, while the black lines show how the time-domain basis size scales vs $dt$ for different pulsars from the $12.5$-year data set.}
    \label{fig:basis_compare}
\end{figure}

In practice, the reduced-rank time-domain basis scales up to higher frequencies (shorter timescales) more efficiently than the classical Fourier basis models. \myfig{fig:basis_compare} illustrates this effect using the pulsars in the $12.5$-year data set -- the time-domain basis with resolution $dt$ always uses a smaller basis size/fewer GP coefficients than the Fourier basis with $N_f = 2/dt$. The improvement in efficiency largely results from the removal of empty columns in $\mathbf{F}$ and therefore depends on the pulsar's data cadence. For some pulsars, the time-domain basis size is an order of magnitude smaller than the Fourier basis size. Given the time to invert the covariance matrix scales as the cube of the basis size, this therefore may net up to $\sim1000$ times more efficient inversions for a given resolution $dt$.

We primarily use this time-domain basis to describe DM and other chromatic ISM effects. As such, the basis matrix, $\mathbf{F}$, will have an additional multiplicative factor of $(\nu/1400\text{ MHz})^{-\chi}$ for each element, where $\chi$ is the chromatic index that takes different values depending on the particular ISM mechanism causing the arrival time variation. For DM, $\chi = 2$ \citep{keith+2013, vanhaasteren+2014_gp}, while for the additional ``chromatic'' kernel, we choose $\chi = 4$ as a proxy for delays that could be induced by scattering \citep{lang71}.

Finally, under the linear interpolation basis, $\phi$ now corresponds to a reduced-rank time-domain covariance matrix. This allows use of a kernel $k(t_k,t_l)$ as a prior over $\phi$, where various kernels may encode different assumptions than the typical Fourier basis + power-law spectral model (\myeq{eq:powerlaw}). Next, we introduce the various kernels we consider for chromatic processes in different pulsars. 

\paragraph{\bf Ridge} Ridge regression can be described as a diagonal kernel with coefficients described by a Gaussian prior:
\begin{eqnarray} \label{eq:ridge}
    k_\mathrm{ridge}(t_k,t_l)
    =& \sigma^2\delta_{kl}
\end{eqnarray} 
where $\sigma$ is an overall variance and $\delta_{kl}$ is the Kronecker delta function between nodes $k$ and $l$ in the linear interpolation basis. In this model, the assumptions are that the delay is uncorrelated epoch-to-epoch and that the amplitude of the delay is finite. This kernel is equivalent to a white noise prior in the Fourier domain, as such it is best suited to model short timescale processes. When used for DM, the ridge kernel is also similar to the standard DMX model as they are both uncorrelated epoch-to-epoch, but the ridge kernel differs because the DMX model places no constraints on the prior variance. As such, the ridge kernel may be less prone to unphysical covariances with other stochastic signals such as chromatic noise due to scattering.

\paragraph{\bf Squared Exponential (SE)}
The SE kernel is well-suited to modeling either short or long timescale variations and is represented as
\begin{eqnarray}
    k_\mathrm{SE}(t_k,t_l) 
    =& \sigma^2\exp{\left(-\frac{|t_k-t_l|^2}{2\ell^2} \right)} + \left(\frac{\sigma}{500}\right)^2\delta_{kl}.
\end{eqnarray}
This introduces the parameter $\ell$, which is a length scale over which the delay will be correlated. The second term is a small regularization factor along the diagonal, which helps ensure stable inversions of the $\phi$ matrix for large values of $\ell$. The $500$ ensures that the regularization is small compared to the value of $\sigma$ and is unitless. One could choose slightly smaller or larger values with very little effect in the models or parameter recoveries. The SE kernel reduces to the Ridge kernel as the parameter $\ell$ tends to zero.

\paragraph{\bf Rational Quadratic (RQ)}
The rational quadratic kernel expands on the squared exponential kernel, and is designed to capture both long and short timescale variations:
\begin{eqnarray}
    k_\mathrm{RQ}(t_k,t_l)
    =& \sigma^2\! \left( 1\! +\! \frac{|t_k-t_l|^2}{2\ell_2^2\alpha_{\rm weight}} \right) ^{-\alpha_{\rm weight}} \!\!\!+ \!\left(\frac{\sigma}{500}\right)^2\delta_{kl} .
\end{eqnarray}
Here, $\alpha_{\rm weight}$ controls the relative weighting of long and short timescale variations. The RQ kernel favors short timescale variations as $\alpha_{\rm weight}$ becomes very large, reducing to the SE kernel with parameters $\sigma$, $\ell_2$. Meanwhile, the RQ kernel approaches a uniform long-timescale variance across all TOAs as $\alpha_{\rm weight}$ tends to zero. This kernel is exclusively used for radio-frequency correlations.

\paragraph{\bf Quasi-periodic (QP)}\label{sec:qp_kernel}
Some DM variations can exhibit periodicity that evolves and changes over time \citep{Madison+2019, jones+2017_ng9_dm}. The quasi-periodic kernel is the combination of a squared exponential kernel (capturing long-timescale variation) and a periodic kernel:
\begin{eqnarray}\label{eq:quasi}
    k_\mathrm{QP}(t_k,t_l) =& \left(\frac{\sigma}{500}\right)^2\delta_{kl} + k_\mathrm{SE}(t_k,t_l)k_\mathrm{P}(t_k,t_l) \nonumber\\
    =& \left(\frac{\sigma}{500}\right)^2\delta_{kl} + \sigma^2 \exp{\left(-\frac{|t_k-t_l|^2}{2\ell^2}    \right)} \nonumber\\ &\times\exp{\left(-\Gamma_p\sin^2{\left(\frac{\pi|t_k-t_l|}{p}\right)}    \right)},\label{eq:qp_kernel}
\end{eqnarray}
where the last line defines $k_\mathrm{P}(t_k,t_l)$, $p$ is the periodicity of variation, $\Gamma_p$ controls the relative weighting of periodicity versus long-timescale variation, and the remaining parameters ($\sigma$, $\ell$) are inherited from the SE kernel. This model of temporal correlation is locally periodic, allowing the shape of the repeating part of the process to change over time. The QP kernel also reduces to the SE kernel as $\Gamma_p$ tends to zero. More details regarding the QP kernel and its relation to the PSD in Fourier space in given in Appendix~\ref{appendix:QP_kernel_PSD}. 

\subsection{Multi-dimensional models}
\label{sec:rf_models}
The previous models all assume there is no $\nu$-dependence, other than an overall scaling proportional to $\nu^{-\chi}$. Here, we introduce kernels that are multi-dimensional along both the time and radio-frequency axes. In general, we can define any kernel as a function of time and radio frequency, $k(t_k,t_l,\nu_m,\nu_n)$; however, it is simpler to use a kernel that is separable so we are able to utilize the kernels we have previously defined:
\begin{equation}
    k(t_k,t_l,\nu_m,\nu_n) = k(t_k,t_l)k(\nu_m,\nu_n).
\end{equation}
In this way, we can attempt to model ``frequency dependent DM'' introduced in \citet{cordes+2016_dm_f}, which arises if multipath propagation induced by $\nu$-dependent refraction causes different radio frequencies of the pulse to sample regions of the ISM with slightly different free electron densities. In \citet{cordes+2016_dm_f}, the kernel is not factorizable, and the radio frequency dependence is non-stationary; however, we can approximate this using a mixture of the aforementioned kernels here. This leads to our final kernel definition.

\paragraph{\bf Quasi-periodic w/ radio correlations (QP\_RF)}
We use a QP kernel for the temporal correlations and an RQ kernel for the radio-band correlations,
\begin{eqnarray}
    k_\mathrm{QP\_RF}(t_k,t_l;\nu_m,\nu_n) &=& 
    k_\mathrm{QP}(t_k,t_l)\; k_\mathrm{RQ}(\nu_m,\nu_n)\nonumber\\
    &&+\left(\frac{\sigma}{500}\right)^2\delta_{kl}\delta_{mn}.
\end{eqnarray}
We can also substitute in the SE kernel for the temporal correlations to create the SE\_RF kernel.

To use this kernel, we must also create a basis in time and radio frequency to allow the multidimensional GP kernel to operate upon. Like the 1D coarse-grained time-domain basis, the 2D time and radio frequency basis is transformed to the full-rank frequency data through interpolation. For simplicity, we switch from a linear interpolation basis to a ``nearest'' interpolation basis, which is more straightforward to implement in multiple dimensions. We first define 3 bands in radio-frequency tailored to the observation ranges for most pulsars in the $12.5$-year data set with edges $\nu' \in [600,1000,1900,3000]$ MHz. This choice nets us a generalized model of ``band noise'' that allows for DM variations in different radio bands to be partially independent. We iterate through these 3 intervals in radio frequency, defining the basis elements as
\begin{align}
    \label{eq:rf_basis}
    F_{ij} = \begin{cases}
        1 &\text{if } t'_{j_t} \le t_i \le t'_{j_t+1} \;\&\; \nu'_{j_\nu} \le \nu_i \le \nu'_{j_\nu+1}, \\
        0 &\text{elsewhere},
    \end{cases}
\end{align}
where $j_t \in [0,\lceil T/dt \rceil]$ iterates through a uniform time grid as previously defined for 1D linear interpolation, $j_\nu \in [0,4]$ iterates through the 3 frequency intervals in $\nu'$, and $j = j_t + j_\nu$. For efficiency, this 2D basis is only computed over the ranges of $\nu'$ at which a particular pulsar has observations; furthermore, nodes are removed for which a given frequency range has no observations in a particular time interval. If neglecting the $\nu$ condition in \myeq{eq:rf_basis}, the basis matrix would reduce to the TOA quantization matrix used to model pulse phase jitter using ECORR parameters. However, as with the time-domain linear interpolation basis, we include an additional $(\nu/1400\text{ MHz})^{-2}$ scaling for each element of the basis such that the basis can model $\nu$-dependent DM. Finally, under this basis we compute an average $t$ and an average $\nu$ in each 2D bin for use as input to the QP\_RF kernel.

\subsection{Deterministic Noise Models}
\label{sec:det_models}

A number of deterministic models can be used to model the perturbations of the interplanetary or interstellar medium along the line of sight to pulsars. The radial motion of the pulsar can induce linearly/quadratically increasing or decreasing DM values \citep{jones+2017_ng9_dm}. These are modeled within the timing model using the first two parameters \texttt{DM1} and \texttt{DM2}, i.e., the quadratic parameters in a polynomial fit for DM variations. Further deterministic models are fit within the \texttt{ENTERPRISE} software package.

\paragraph{\bf Annual variation (AV)}
Annual DM variations are historically fit across pulsars and are thought to be due to the annual periodic sampling of slightly different paths through the ISM due to the Earth's motion around the Sun \citep{Madison+2019}. The model is 
\begin{equation}
    \Delta t_{\rm AV} = A \sin\left(2\pi f_{\rm yr} t + \phi \right)\left(\frac{\nu}{\nu_{\rm ref}}\right)^{-2},
\end{equation}
where $A$ is the amplitude of the sinusoidal variations and $\phi$ is a variable phase term. 

\paragraph{\bf Transient Chromatic Events}
Some pulsars show dramatic $\nu$-dependent changes in pulse times of arrival over short timescales. These have often been attributed to transient ISM under/over-densities or scintillation off of such structures \citep{lam_j1713_2nd_2018}. We use two different transient models. The first is an exponential dip model, which has a sudden drop followed by an exponential recovery:
\begin{equation}
    \label{eq:dip}
    \Delta t_{\rm dip} = A \Theta(t-t_0)e^{- \frac{t-t_0}{\tau} }\left(\frac{\nu}{\nu_{\rm ref}}\right)^{-2},
\end{equation}
where $\tau$ is the rate of recovery, $A$ is the amplitude dip, and $t_0$ is the time of the event. This first model was derived phenomenologically to explain the ``events" in the DM time series in \psrplus{J1713}{0747} \citep{lam+2018_j1713_2nd}.
The second, more general, model for dramatic short timescale variations is a cusp model\footnote{Many of the cusps in various pulsars seem attributable to the SW.}:
\begin{align}
    \label{eq:cusp}
    \Delta t_{\rm cusp} &= A \left[ \left[1 - \Theta(t-t_0) \right] e^{- \frac{t_0 - t}{\tau_{\rm pre}}}\right.\nonumber \\ 
    &\quad\quad\quad+ \left.\Theta(t-t_0) e^{- \frac{t - t_0}{\tau_{\rm post}}}\right]
    \left(\frac{\nu}{\nu_{\rm ref}}\right)^{-\chi} ,
\end{align}
where $A$ is the amplitude of the deviation, which can be either positive or negative, $t_0$ is the time of the event, $\tau_{\rm pre}$ is the rate of exponential change prior to the time $t_0$, and $\tau_{\rm post}$ is the rate of exponential recovery following the time $t_0$. The dip model (\myeq{eq:dip}) is a specific sub-model of the general cusp model (\myeq{eq:cusp}), where $\tau_{\rm pre} = 0$ and $A < 0$, however, due to its historical significance, we include it as a stand-alone model in this work.  To add maximal flexibility to this model, we also allow for a variable $\nu$-frequency dependent scaling during these deviations,
$\Delta t_{\rm cusp} \propto (\nu/1400\text{ MHz})^{-\chi}$.

\paragraph{\bf Solar Wind (SW)}
The spectral character of the solar wind is distinct from that of a changing ISM \citep{tiburzi+2016,tiburzi+2021,Madison+2019}, which makes it useful to model the solar wind as a separate dispersive signal in pulsar data sets. The solar wind model used here is based on the work in \citet{hazboun+2022sw}, which offers a variety of schema for dispersive solar wind modeling. Here we use the binned approach studied from \citet{hazboun+2022sw}, which fits a constant value for $n_E$ across $6$-month-long bins in all pulsars independently. As noted in \mytab{tab:priors}, we use a uniform prior in the solar electron number density, $n_E$, rather than the astrophysical prior derived from in-situ satellite measurements. Uniform priors allow us to more easily leverage factorized likelihood methods \citep{taylor+2022_fl}, i.e., taking the product of posteriors for $n_E$ across all pulsars. A $6$-month bin was chosen such that it is narrow enough that the binned SW values can be set constant across the pulsars in a final noise analysis. We note that recent results demonstrate that PTA data sets benefit from more time-variable solar wind modeling \citep{nitu+2024, susurla+2024}. Future work explores the incorporation of time-domain GP models for a time-variable solar wind (Larsen, et al., In Prep). More about how the factorized SW model is used here across the individual pulsars is discussed in \mysec{sec:methods}. Following \citet{hazboun+2022sw}, we fit for a higher-order term in the SW model, $n_E^{(4.39)}$, which does not vary in time and allows a more flexible model than the simplest $1/R^2$ fall off used to model SW electron density. This is also studied in a factorized-likelihood framework and is highly favored by the data in $2$ particular pulsars. See \mysec{sec:results}.

\section{Methods}
\label{sec:methods}

To select optimal chromatic models for our 20 pulsar subset of the NANOGrav $12.5$-year data set, and understand the effects of these chromatic models on the overall characterization of each pulsar, we apply Bayesian model selection and parameter estimation techniques. We apply these methods very similarly as done in previous works (e.g., \citealt{lentati+:2016,goncharov+2021gwb, ng15detchar, Miles+2025_noise}), with implementations following especially closely to \citet{ng15pipeline} where further details may be found.

To summarize, we use the \texttt{enterprise} \citep{enterprise} and \texttt{enterprise\_extensions} \citep{enterprise_ext} software packages to construct our model likelihood $\mathcal{L}(\vec{\delta t}|\vec{\eta},\mathcal{M})$ and priors $\pi(\vec{\eta}|\mathcal{M})$ given model $\mathcal{M}$ and hyperparameter vector $\vec{\eta}$, where the goal is typically the estimation of $\vec{\eta}$ under the posterior probability $\mathcal{P}(\vec{\eta}|\vec{\delta t},\mathcal{M}) \propto \mathcal{L}(\vec{\delta t}|\vec{\eta},\mathcal{M})\pi(\vec{\eta}|\mathcal{M})$. Our likelihood is a multivariate Gaussian $\mathcal{L}(\vec{\delta t}|\vec{\eta},\mathcal{M}) \sim \mathcal{N}(0, \mathbf{C})$, where the full-rank timing residual covariance matrix $\mathbf{C} = \mathbf{N} + \mathbf{T}\mathbf{B}\mathbf{T}^T$ is composed of a white noise matrix $\mathbf{N}$ depending on per-receiver/backend EFAC, EQUAD, and ECORR noise parameters, while $\mathbf{T}$, $\mathbf{B}$ are respectively the concatenation of all reduced-rank GP design matrices and prior variance matrices, which includes the linear timing model, achromatic RN, and our new chromatic GPs from \mysec{sec:models}. Notably, the timing models in this work are equivalent to those from \citet{arz:2020}, except all DMX parameters have been removed and replaced with 2 parameters allowing for a quadratic trend in DM (\mysec{sec:det_models}).

Critical in this work is the model selection. We compute a Bayes factor under various model hypotheses, given as the ratio of marginal likelihoods (aka model evidences) under two models $\mathcal{M}_1$ and $\mathcal{M}_2$,
\begin{align}
    \mathcal{B}^{\mathcal{M}_2}_{\mathcal{M}_1} = \frac{\mathcal{Z}(\vec{\delta t}|\mathcal{M}_2)}{\mathcal{Z}(\vec{\delta t}|\mathcal{M}_1)}.
\end{align}
where the marginal likelihood $\mathcal{Z}$ is typically found via integration over the multidimensional prior volume $\Omega$,
\begin{align}
    \mathcal{Z}(\vec{\delta t}|\mathcal{M}) = \int_\Omega\mathcal{L}(\vec{\eta}|\vec{\delta t},\mathcal{M})\pi(\vec{\eta}|\mathcal{M})d\vec{\eta}.
\end{align}
The Bayes factor is equivalent to a ratio giving the odds of a hypothesis given by $\mathcal{M}_2$ vs the hypothesis given by $\mathcal{M}_1$ as informed by the data (assuming equal \emph{prior} odds), and may be used to select amongst appropriate models. Namely, $\mathcal{B}^{\mathcal{M}_2}_{\mathcal{M}_1} > 1$ favors use of $\mathcal{M}_2$, whereas $\mathcal{B}^{\mathcal{M}_2}_{\mathcal{M}_1} < 1$ favors use of $\mathcal{M}_1$. Instead of direct estimation of the model evidence (e.g., using nested sampling), we estimate Bayes factors using product space sampling \citep{Lodewyckx+2011, hhh+2016}, implemented as the \texttt{HyperModel} in \texttt{enterprise\_extensions}, in which two or more models are sampled together with a ``switch'' hyperparameter, the posterior of which is used to directly estimate an odds ratio \citep{ng15pipeline}. This frames the model selection as a parameter estimation problem, which can be evaluated using Markov Chain Monte Carlo (MCMC) sampling. In the case of \emph{nested models}, where $\mathcal{M}_2$ reduces to $\mathcal{M}_1$ under some particular combination of hyperparameters $\vec{\eta}_0$ in $\mathcal{M}_2$, the Bayes factor is more easily approximated as the prior-to-posterior density ratio at the location of the parameters $\vec{\eta}_0$ (Savage-Dickey ratio; \citealt{Dickey1971}),
\begin{align}
    \mathcal{B}^{\mathcal{M}_2}_{\mathcal{M}_1} = \frac{\pi(\vec{\eta} = \vec{\eta}_0|\mathcal{M}_2)}{\mathcal{P}(\vec{\eta} = \vec{\eta}_0|\vec{\delta t},\mathcal{M}_2)}.
\end{align}
For example, a Bayes factor on a model \emph{with} achromatic red noise vs the same model \emph{without} achromatic red noise is inversely proportional to the posterior in the tail $\mathcal{P}(\log_{10}A_{\mathrm{RN}}=-18|\vec{\delta t},\mathcal{M})$. Both the model selection and final estimation of noise parameter values are carried out using \texttt{PTMCMCSampler} \citep{ptmcmcsampler}, with further details on the sampler's implementation presented in \citet{ng15pipeline}.

We note that recent works \citep{vH2025_model_averaging, dimarco+2025} advocate for approaches using model \emph{averaging} as opposed to model \emph{selection}, wherein \emph{all} expected noise processes be represented under the model with suitable priors to allow the model to drop in or out as needed. We stick to the classical approach, noting that averaging among multiple \emph{different GP noise models} for the \emph{same physical process} still requires more efficient implementations of transdimensional MCMC \citep{EllisCornish2016} to be tractable in full-PTA analyses.

\subsection{Model Customization Framework}

\begin{figure}
    \centering
    \includegraphics[width=0.45\textwidth]{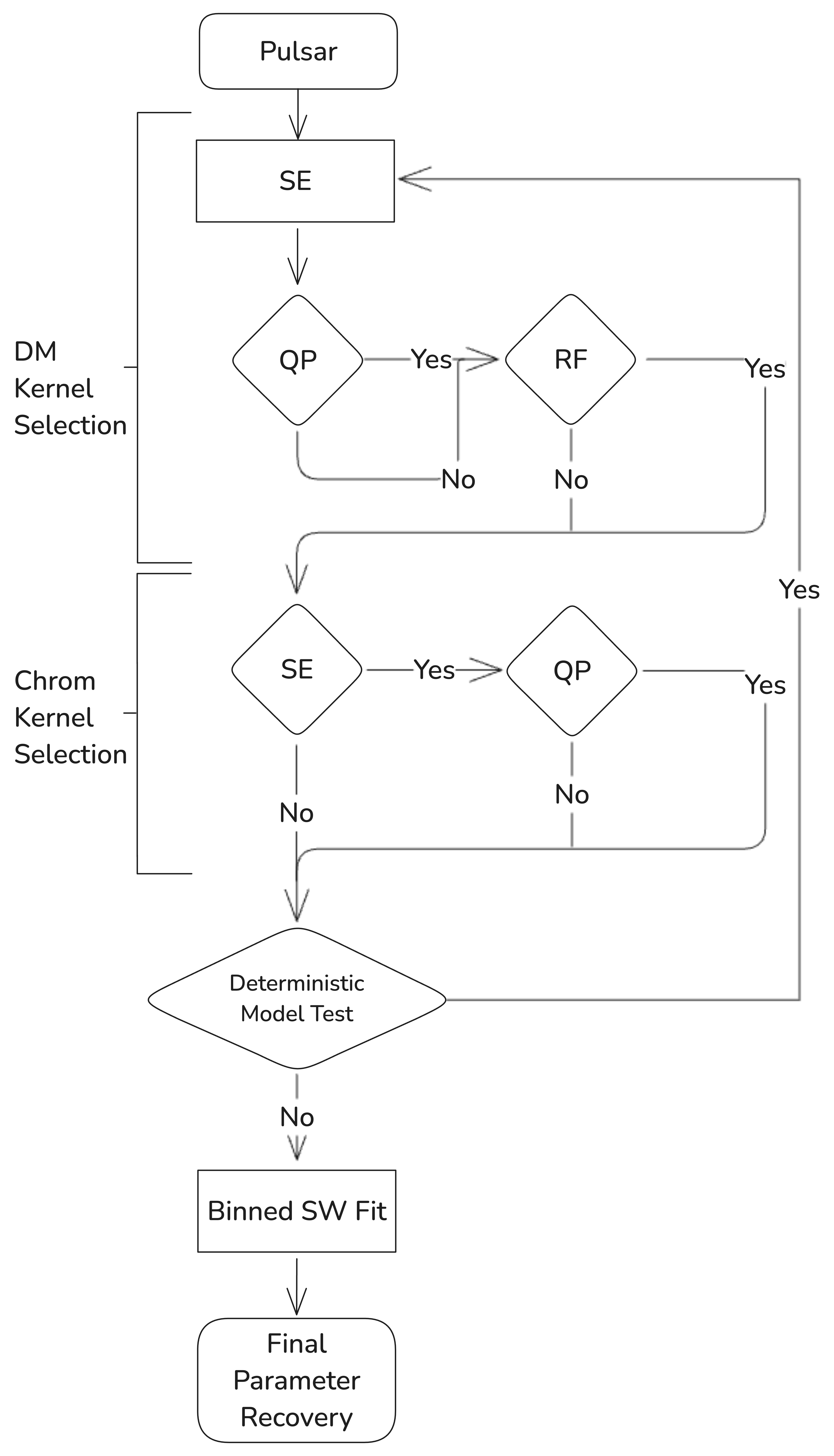}
    \caption{Flowchart illustrating the full model customization framework. For each pulsar, we begin by selecting the DM kernel, Initially set to the SE kernel for all pulsars. We then check if the pulsar additionally favors a QP kernel, and then check the radio frequency dependence. Next we select the Chrom kernel: If the pulsar favors an SE kernel, we test for QP, otherwise no Chrom kernel is applied. After kernel selection we check for deterministic signals. If one is found we repeat the kernel selection process. Once no further deterministic signals are detected, we bin the SW and fix the values to the median across all pulsars. Final parameter estimation is then performed using these fixed SW densities.}
    \label{fig:ModelFlowchart}
\end{figure}

The customization of each pulsar's chromatic model is carried out in multiple phases, as evaluating various components of the model requires distinct analyses. Namely, the relative odds ratios of the models is an important part of deciding which model to use, but is only one in a rubric used to choose a bespoke model for each of these pulsars. These phases are:
\begin{enumerate}
    \item \textbf{Kernel Selection:} We start by selecting which set of time-domain kernels is favored by the data for each pulsar's DM and chromatic noise. Under the \texttt{HyperModel}, we test 4 DM kernels (SE, QP, SE\_RF, and QP\_RF) and 3 options for chromatic kernels (SE, QP, or no kernel), the combinations of which make for 12 possible models. This evaluation is split across multiple \texttt{HyperModel} analyses, with a preliminary analysis to test the evidence of DM and chromatic noise, and at least one additional to refine amongst the remaining kernel options. Whichever model has the largest number of samples based on the \texttt{HyperModel} switch parameter is selected for the next phase of analysis. During this first phase, a single time-independent solar wind parameter $n_E=6.9\;{\rm cm}^{-3}$ \citep{hazboun+2022sw} is sampled across the whole data set. \psrplus{J1713}{0747} also began the analysis with 2 deterministic exponential dip models whose presence were known \emph{a priori} from \citet{lam+2018}.  
    \item \textbf{Assess the Kernel \& Transient Components:} Once the model selection is performed, we perform a separate parameter estimation to accrue more samples and evaluate the model. To identify various possible transient or deterministic components discussed in \mysec{sec:det_models}, we use the \texttt{la\_forge} software to create time-domain reconstructions showing the effects each GP signal would induce on the timing residuals. If there is any apparent non-stationarity or annual trend in the signal, we perform another \texttt{HyperModel} analysis with one or more deterministic signals included in the model, establishing evidence for or against their inclusion. Additionally, inspection of the posterior DM parameters suggested further changes to the DM kernels for 2 pulsars: \psrplus{B1937}{21} whose preferred model includes 2 DMGPs (\mysec{sec:B1937}), and \psrminus{J1455}{3330} whose preferred model uses the simplest Ridge kernel (\mysec{sec:J1455}).
    \item \textbf{Binned SW Fit:} With appropriate chromatic models selected, we next perform a parameter estimation run for each pulsar where the SW density is allowed to vary with a separate parameter for each 6-month interval as well as the higher-order SW term (\mysec{sec:det_models}). The factorized likelihood product of the SW density in each bin is then taken, with the values fixed to the median in each bin for subsequent analyses.
    \item \textbf{Reanalysis \& Parameter Estimation:} We end by performing at least one more parameter estimation run using the fixed, binned, PTA-wide SW densities. We use the Savage-Dickey Bayes Factor to assess if previously significant model components, such as the more complex kernel parameters and additional transient terms, are still needed under the updated SW model. If they are no longer favored, we remove them from the model.
\end{enumerate}
This process is repeated for every pulsar in the data set, with each phase performed in parallel. Throughout every phase the constituent pieces of the timing model, white noise, and achromatic red noise models are left unchanged, but their parameters are varied as part of the analyses. 

\section{Results}
\label{sec:results}

\input{Table_ModelResults}

The chosen noise model for each pulsar can be found in \mytab{tab:modelresults}. Based on the average observing cadence, the standard basis size for the time-domain models ($dt = 15$ days) was used for $16$ of the $20$ pulsars. Two pulsars (\psrplus{J0645}{5158} and \psrminus{J1455}{3330}) were run with a $dt=7$, while the remaining two (\psrplus{B1937}{21} and \psrplus{J1713}{0747}) required a much finer basis size ($dt = 3$) to appropriately capture chromatic variations, as these were observed at both AO and the GBT. These two pulsars both have a history of unique noise characteristics \citep{vivekanand2020, goncharov+2020}, and their specific noise model preferences will be discussed in detail later. The effect of changing the basis size is further explored in \S\ref{sec:basis_size}.

There is no clear preference for a specific kernel with most pulsars split between either the SE or QP models; however two pulsars prefer a radio-frequency dependent DM kernel (\psrminus{J1600}{3053} and \psrplus{J1713}{0747}) and \psrminus{J1455}{3330} prefers the simplest model, Ridge (\myeq{eq:ridge}). 
There are $14$ pulsars that favor including an additional chromatic process with a steeper frequency dependence than DM ($\nu^{-4}$), consistent with what is seen in other PTA data sets \citep{eptadr2_2:noise, pptadr3:noise, Miles+2025_noise}. However, as in the case of the DM kernels, there is again no clear preference for a specific kernel. The kernel parameter values and credible intervals are shown in \mytab{tab:chrom_params}. 

Additional deterministic signals were preferred in six pulsars with the majority of additional signals being a non-stationary chromatic event best modeled as a cusp (\myeq{eq:cusp}). There are two pulsars (\psrminus{J0613}{0200} and \psrminus{J1614}{2230}) that prefer an additional annual variation (AV) signal. The deterministic parameter values and credible intervals are shown in \mytab{tab:det_params}. Both of these pulsars were found to have significant (close to) annual trends in \citet{jones+2017_ng9_dm}.

\begin{figure}
    \centering
    \includegraphics[width=0.45\textwidth]{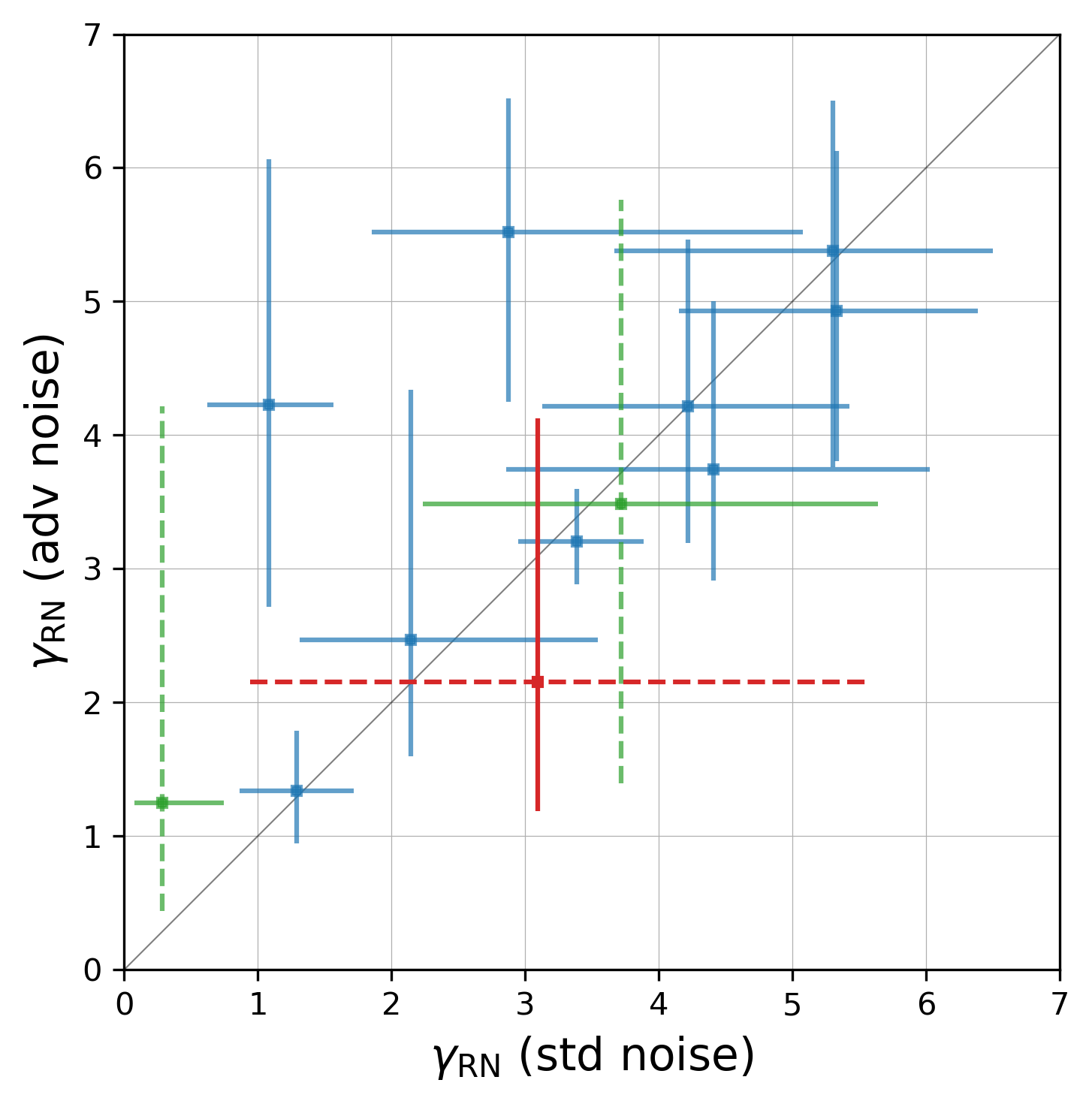}
    \caption{Red noise spectral index ($\gamma_{\rm RN}$) posterior changes between NANOGrav's standard noise model and the ``customized" model found in this work. The points represent the median posterior values while the crosses cover the 16$^{\rm th}$ to 84$^{\rm th}$ percentile regions. The blue markers indicate pulsars where red noise was found to be present in both cases. The green markers indicate pulsars (\psrminus{J1600}{3053} and \psrplus{J2043}{1711}) where red noise was found to be significant under the standard noise model, but insignificant under custom noise models, while the red marker indicates a pulsar (\psrminus{J1455}{3330}) where the insignificant red noise recovered with the standard noise model was found to be significant when utilizing a custom noise model. In general, shallow spectral indices indicate that unmodeled chromatic noise may be leaking into the red noise model, while steeper spectral indices are more in line with the expectations for red noise, either from individual pulsar mechanisms, such as spin noise, or from a common GWB. There are eight pulsars where there is no significant red noise found when using either model; as such, they are not shown in this plot.}
    \label{fig:RNcompare}
\end{figure}

At a glance, it is difficult to draw any global conclusions from the inclusion of these new models. It has been well documented that excess, and/or mismodeled, noise is not restricted to an individual pulsar's noise model -- thus, these noises may pollute a search for a gravitational wave background \citep{dimarco+2025}. As such, it can be difficult to interpret changes in noise parameter posteriors and how they might affect a GWB search. However, when excess chromatic noise or excess white noise is present, it tends to lower the pulsar's achromatic red noise model spectral index \citep{dimarco+2025, ferranti+2025}. Meanwhile, an astrophysical GWB is expected to be a steeper RN process with $\gamma = \frac{13}{3}$ \citep{Phinney2001}. Thus, we have a general sense that a steeper achromatic RN power law tends to mean that a pulsar's noise model has better isolated white and/or chromatic noise, which in turn increases the pulsar's sensitivity to the GWB \citep{Miles+2025_noise, dimarco+2025, ferranti+2025}.

In \myfig{fig:RNcompare}, the achromatic RN spectral index ($\gamma_{\mathrm{RN}}$) is plotted for the standard noise model as well as the preferred custom noise model for the 12 pulsars where significant RN was found using either one or both of the models. The other eight pulsars did not have significant RN under either model. Overall, there is a slight trend from shallower spectral indices with the standard models towards steeper spectral indices under the preferred, custom models used in this work. In two cases (\psrminus{J1600}{3053} and \psrplus{J2043}{1711}), the preferred model no longer finds significant RN that had been present under the standard model. While in one case (\psrminus{J1455}{3053}), insignificant RN in the standard noise model was found to be significant under the preferred model. 
Each of these cases will be discussed in more detail in \S\ref{subsec:ind_pulsars}. The achromatic RN parameter medians, credible intervals, and/or upper limits under both modeling assumptions can be found in \mytab{tab:rn_params}.

\begin{figure*}
    \centering
    \includegraphics[width=0.9\textwidth]{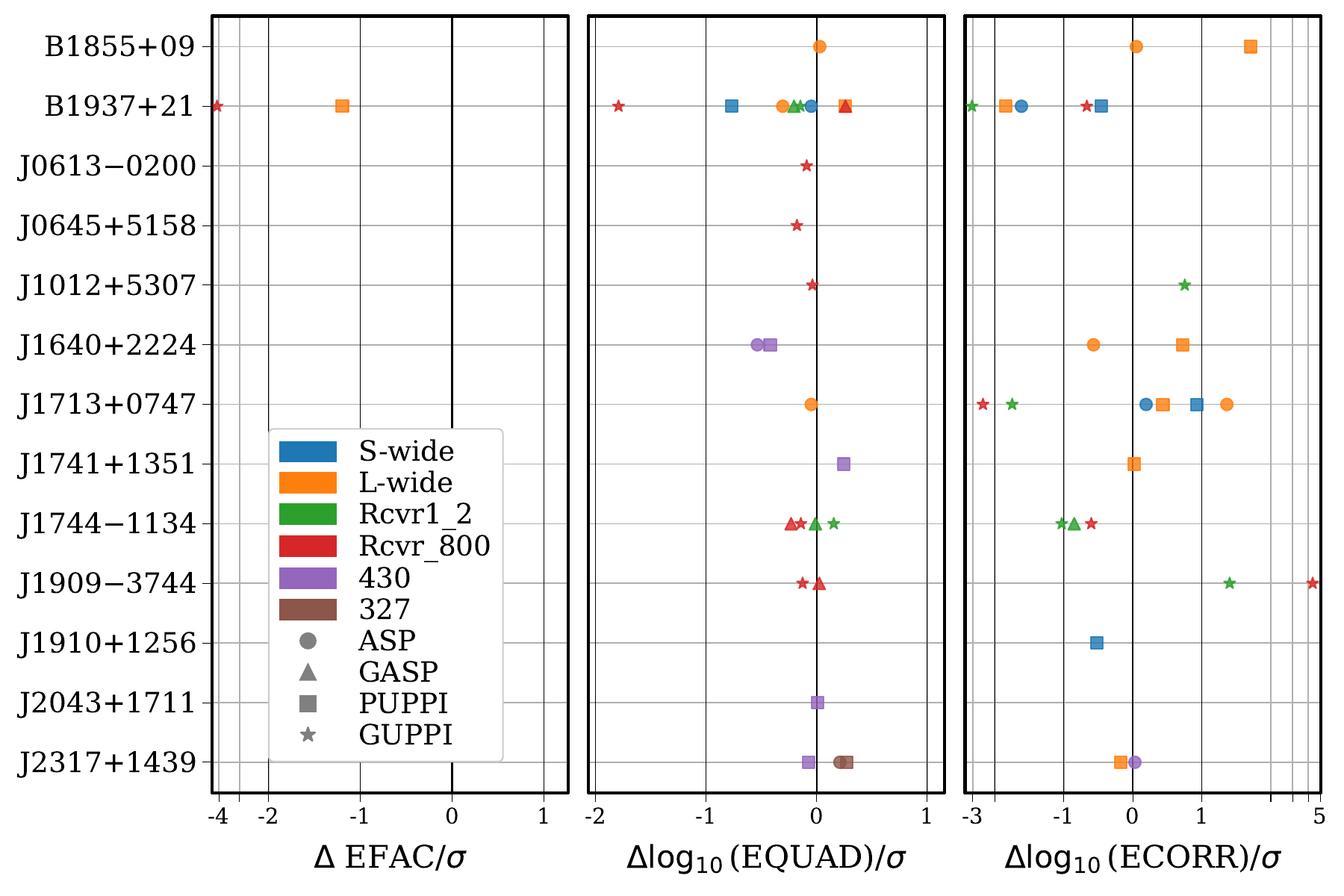}
    \caption{White noise changes between NANOGrav's standard noise model and the customized noise model found in this work. We only plot parameters which are significant in both the standard noise and custom noise. All plots show the difference in median posterior value relative to the $1\sigma$ uncertainty from the standard noise posteriors. Negative values show where the new models from this work reduce the white noise values, while positive values show an increase. The different colors and markers correspond to the various receiver and backend combinations present in the data. Differences in EFAC are shown on the left, EQUAD in the center, and ECORR on the right. The changes to EFAC and EQUAD are relatively minor, however, \psrplus{B1937}{21} displays a significant reduction in both EFAC and EQUAD from the standard model to chromatic model. The ECORR changes are much more dramatic across the board, with no clear trends visible in this plot, however, we will discuss the significance of these changes and some potential causes in \S\ref{sec:chrom_models_ecorr}. \label{fig:WNcompare}}
\end{figure*}

\begin{figure*}
    \centering
    \includegraphics[width=0.9\textwidth]{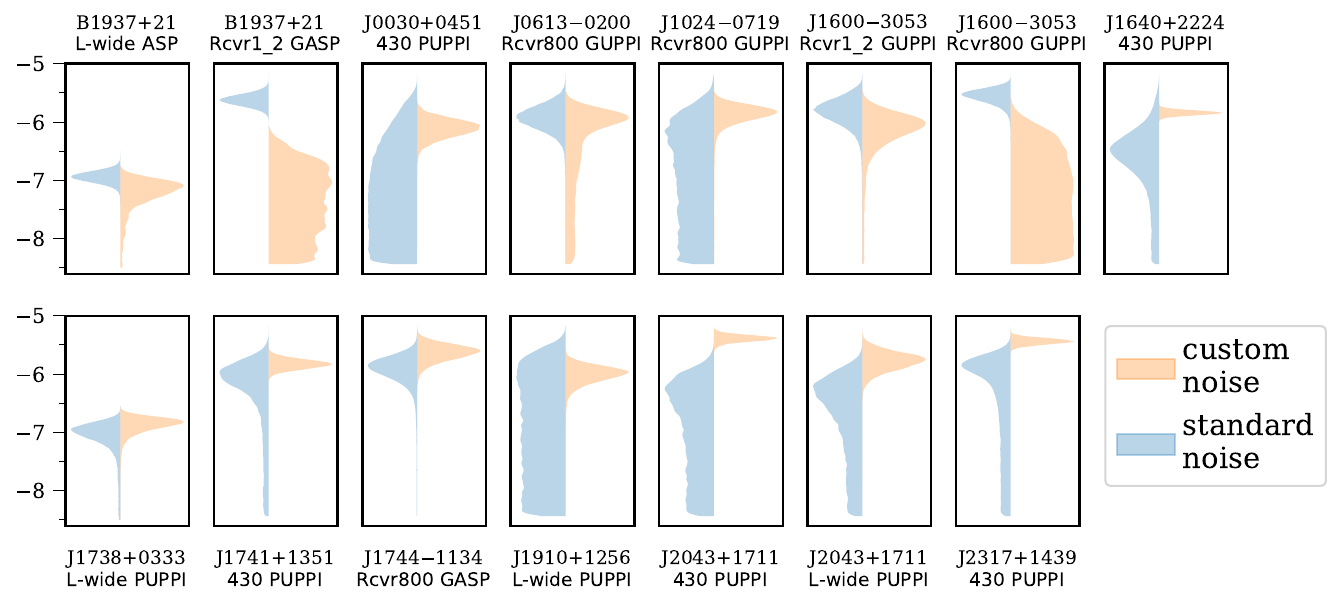}
    \caption{Pulsar/backend combinations that showed changes in ECORR significance with custom noise modeling, corresponding to the bolded entries in \mytab{tab:ecorr_params}. \label{fig:ecorr_posteriors}}
\end{figure*}

In \myfig{fig:WNcompare} the WN changes between the standard noise model and the preferred custom noise model found in this work are shown for WN parameters that are significant under both models. In NANOGrav's modeling approach, there are three WN terms: EFAC, EQUAD, and ECORR. There is a unique parameter value of the WN for each telescope receiver and backend combination, which are represented by different colors and marker combinations, respectively, as explained in the legend of \myfig{fig:WNcompare}. Similarly to what was found in the RN comparison above, there is no significant change for many pulsars. Specifically for EFAC and EQUAD, there are minimal changes outside of \psrplus{B1937}{21},
which sees significant reductions in WN values (\psrplus{J1713}{0747} also experiences reductions in EFAC/EQUAD where the parameter values are no longer significant under the custom noise model). The most distinct changes are found when assessing the ECORR parameter, with several ECORRs changing in significance as shown in \myfig{fig:ecorr_posteriors}. While there remains no definitive trend, with the known interactions between ECORR, DMX, and achromatic RN \citep{hazboun:2020slice}, it should not be surprising to see ECORR models changing significantly in the presence of these new noise models. See \mysec{sec:chrom_models_ecorr} for further discussion about the interactions between the chromatic models and ECORR.

\subsection{Solar Wind} 

\begin{figure}
    \centering
    \includegraphics[width=0.45\textwidth]{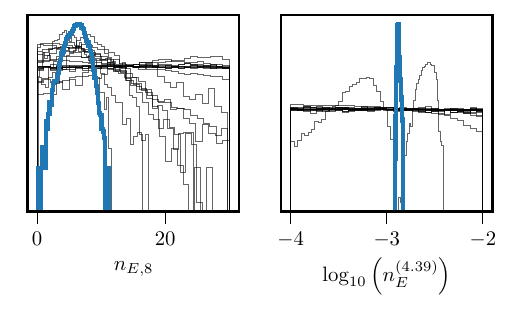}
    \caption{Posterior distributions for two representative solar wind electron density parameters. \emph{Left panel:} The posteriors for the 8th solar wind bin's electron density across all of the pulsars (gray) and the product of the posteriors. \emph{Right panel:} The posteriors for the coefficient of the higher-order correction of the solar wind electron density model. There are only two significant posteriors. The one centered at $\log_{10}n_E^{(4.39)}\approx-2.5$ is from \psrplus{J0030}{0451} while the one centered near $\log_{10}n_E^{(4.39)}\approx-3.2$ is for \psrminus{J1614}{2230}.}
    \label{fig:sw_posteriors}
\end{figure}

Of the $25$ solar wind bins in the data set, we show the $8$th bin in \myfig{fig:sw_posteriors} as a representative sample. The gray contours are the posteriors of $n_E$ over that particular bin measured by individual pulsar lines of sight past the Sun which spans MJDs $54725-54913$. A small fraction of the pulsars are uninformative here, but most are mildly peaked around $6$ cm$^{-3}$ or are ``upper-limits'' on $n_E$. The blue contours represent the product of all the posteriors of the $20$ pulsars for the left and right figures.

The higher-order solar wind term here is taken to be constant across the entire data set and is most pronounced at cusps near minimum solar elongation. Here we show that the higher-order term is only significant in \psrplus{J0030}{0451} and \psrminus{J1614}{2230}, which coincides with measurements taken of both of the pulsars at very low solar-elongation. See \citet{hazboun+2022sw} Figure $4$. While their posteriors are consistent with each other, they are peaked at different values. The resulting posterior product among the pulsars ends up being well peaked right between the individual peaks.

\subsection{Individual Pulsars' Preferred Models}
\label{subsec:ind_pulsars}

As mentioned in \mysec{sec:intro}, the 20 pulsars in this analysis were chosen for myriad reasons, e.g, their total time span of data or high levels of suspected chromatic noise. One of the most important conclusions we come to through this analysis is that each pulsar prefers a noise model different in some regard from the others, which can be explained by the fact that pulsars have unique lines of sight through the IPM/ISM. Here we go through each pulsar in the analysis and highlight interesting details of their custom chromatic noise models. 

Throughout this section, we will use the term \textit{``change in significance"} when going from an upper limit to a credible interval or vice versa (reaches a $> 10$ Bayes factor threshold). We will use the term \textit{``significant change"} when referring to a ($> 1\sigma$ change) in credible interval.

\subsubsection{\psrplus{J0030}{0451}}
The chromatic model \psrplus{J0030}{0451} included a significant DMGP with a QP kernel and a significant Chrom GP with a QP kernel. This pulsar is close to the ecliptic, and as shown in \myfig{fig:sw_posteriors}, is one of two pulsars that favors a higher-order term in the SW model. 

The WN in this pulsar is featured in \myfig{fig:ecorr_posteriors} as it has a change in significance at 430 MHz with the PUPPI backend, with the rest remaining insignificant. This newly significant ECORR supports the possible need for a more sophisticated SW model. The RN has no significant change. 

\subsubsection{\psrminus{J0613}{0200}}
The chromatic model for \psrminus{J0613}{0200} included a significant DMGP with a QP kernel, and no significant Chrom GP. This pulsar appears in \mytab{tab:det_params} due to having significance for deterministic signals from AV and chromatic ($\nu^{-4}$) dip.

\psrminus{J0613}{0200} was studied with the EPTA DR2 and and they did not observe strong evidence for an AV signal \cite{Chalumeau+2022}. The MPTA does find a significant AV signal for \psrminus{J0613}{0200} in \citep{mpta-dr2:noise}.

The WN in this pulsar is featured in \myfig{fig:ecorr_posteriors} as it has a change in significance at 800 MHz with the GUPPI backend, with the rest remaining insignificant. The RN has no significant change. 

In \citet{Larsen+2024}, it was found using the NANOGrav 15-year data set that the additional chromatic model is favored for \psrminus{J0613}{0200}, and this results in a steeper achromatic RN spectrum than using standard DMX. However, here the additional chromatic GP is not significantly favored. This suggests that longer timespan data sets are beneficial both to tease out these higher-order chromatic effects and improve the characterization of achromatic RN.

\subsubsection{\psrplus{J0645}{5158}}
The chromatic model for \psrplus{J0645}{5158} included a significant DMGP with a SE kernel and a significant Chrom GP with a QP kernel. This pulsar also has an ecliptic latitude $<30^\circ$, which means that this pulsar may need a more sophisticated SW model than the one presented in this paper. This pulsar required a finer basis size $dt=7$ compared to the majority of pulsars that had fixed $dt=15$.

The WN and RN in this pulsar had no significant changes.

\subsubsection{\psrplus{J1012}{5307}}
The chromatic model for \psrplus{J1012}{5307} has a significant DMGP with a QP kernel, and a significant Chrom GP with a SE kernel.

The WN in this pulsar had no significant change. The RN had a significant decrease in the amplitude, but no change in spectral index, despite introducing the new chromatic model. This pulsar was studied further in \citep{Larsen+2024}, where this effect was also noted in the NANOGrav 15-yr data set, and it was found that the achromatic RN displayed anticorrelations with the chromatic noise over time.

\subsubsection{\psrminus{J1024}{0719}}
The chromatic model for \psrminus{J1024}{0719} has a significant DMGP with a SE kernel and no significant Chrom GP. This pulsar was analyzed with multiple basis sizes for $dt$ to explain higher ECORR values, the results of which are discussed in \mysec{sec:basis_size}. In the analysis with this pulsar, none of the ECORR values changed appreciably with the basis size.

The WN in this pulsar is featured in \myfig{fig:ecorr_posteriors} as it has a change in significance at 800 MHz with the GUPPI backend. The RN did not change significantly

\subsubsection{\psrminus{J1455}{3330}}
\label{sec:J1455}
The chromatic model for \psrminus{J1455}{3330} has a significant DMGP with a Ridge kernel, and no significant Chrom GP. This pulsar used a finer basis size $dt=7$ compared to the majority of pulsars that had fixed $dt=15$. \psrminus{J1455}{3330} is studied further in \citet{ng15_j1455_dm2025_lam+} for the impact of simplified DM modeling on gravitational wave sensitivity.

The WN in this pulsar remains insignificant. The RN in this pulsar is highlighted in \myfig{fig:RNcompare} as the previously insignificant red noise from the standard noise model is now significantly measured with the custom noise model.

\subsubsection{\psrminus{J1600}{3053}}
The chromatic model for \psrminus{J1600}{3053} has a significant DMGP with a QP\_RF kernel and a significant Chrom GP with a QP kernel.

The WN in this pulsar is featured in \myfig{fig:ecorr_posteriors} as it has a change in significance at 800 MHz and 1.2 GHz with the GUPPI backend. The RN in this pulsar is highlighted in \myfig{fig:RNcompare} as the significant red noise from the standard noise model had a change in significance with the custom noise model, making it insignificant.

This pulsar is studied further in \citet{Larsen+2024}, where it was found that applying a
custom Chrom GP resulted in a steeper achromatic red
noise spectrum than using standard DMX, which differs from our result where the red noise became insignificant with the custom model.

\subsubsection{\psrminus{J1614}{2230}}
The chromatic model for \psrminus{J1614}{2230} has a significant DMGP with a QP kernel, a significant Chrom GP with a SE kernel. This pulsar appears in \mytab{tab:det_params} due to having significance for deterministic signals from annual DM variations. This pulsar also has an ecliptic latitude $<30^\circ$, which means that this pulsar may need a more sophisticated SW model than the one presented in this paper. In \myfig{fig:sw_posteriors}, this pulsar is one of two pulsars that favor a higher-order term in the SW model.

The WN and RN in this pulsar have no significant changes.

\subsubsection{\psrplus{J1640}{2224}}
The chromatic model for \psrplus{J1640}{2224} has a significant DMGP with a QP kernel and a significant Chrom GP with a SE kernel. 

The WN in this pulsar is featured in \myfig{fig:ecorr_posteriors} as it has a change in significance at 430 MHz with the PUPPI backend. The RN has no significant change.

\subsubsection{\psrplus{J1713}{0747}}
\label{sec:J1713}
The chromatic model for \psrplus{J1713}{0747} has a significant DMGP with a QP\_RF kernel and a significant Chrom GP with a QP kernel. This pulsar appears in \mytab{tab:det_params} due to having significance for deterministic signals from a DM dip ($\nu^{-2}$), and chromatic dip ($\nu^{-1.35}$), where the second dip's chromatic index is fit for as it is known to depart from a purely dispersive trend \citep{goncharov+2020, Chalumeau+2022, Larsen+2024}. This pulsar required a finer basis size $dt=3$ compared to the majority of pulsars that had fixed $dt=15$.

This pulsar has been extensively studied by NANOGrav and other PTAs. In particular, due to recent chromatic timing events resulting in pulse shape changes \citep{lam_j1713_2nd_2018, goncharov+2020, atel14642_J1713dip, atel14652_J1713dip, lam_j1713_3rd_2021, lin_j1713_3rd_2021, singha_j1713_3rd_2021, jennings_j1713_3rd_2024, mandow_J1713_2025}. This pulsar is also studied further with Chrom GP noise models in \citep{Larsen+2024}, where it was found that using a custom Chrom GP model over a DMX model resulted in a significant change in achromatic RN parameters. Additionally, the posterior parameters for the chromatic dips found using the 15-year data set (\citealt{Larsen+2024}; Figure 7) are consistent with those found here.

The WN in this pulsar has a significant change with the L-wide ASP and 800 MHz with the GUPPI backend. This pulsar appears in \myfig{fig:WNcompare} in which significant reductions in EFAC, EQUAD, and ECORR are shown. The RN has a significant decrease in amplitude and a significant increase in spectral index, bringing it more in line with the common process. 

\subsubsection{\psrplus{J1738}{0333}}
The chromatic model for \psrplus{J1738}{0333} has a significant DMGP with a SE kernel, and a significant Chrom GP with a SE kernel. This pulsar also has an ecliptic latitude $<30^\circ$, which means that this pulsar may need a more sophisticated SW model than the one presented in this paper.

The WN in this pulsar is featured in \myfig{fig:ecorr_posteriors} as it has a change in significance with the L-wide PUPPI backend. The RN has no significant change.

\subsubsection{\psrplus{J1741}{1351}}
The chromatic model for \psrplus{J1741}{1351} has a significant DMGP with a SE kernel, and no significant Chrom GP. 

The WN in this pulsar is featured in \myfig{fig:ecorr_posteriors} as it has a change in significance at 430 MHz with the PUPPI backend. This pulsar was analyzed with multiple basis sizes for $dt$ to explain higher ECORR values, the results of which are discussed in \mysec{sec:basis_size}. In the analysis with this pulsar, none of the ECORR values changed appreciably with the basis size. The RN has no significant change.

\subsubsection{\psrminus{J1744}{1134}}
The chromatic model for \psrminus{J1744}{1134} has a significant DMGP with a SE kernel and a significant Chrom GP with a SE kernel. This pulsar is studied further in \citet{Larsen+2024}, where it was found that achromatic RN was no longer significant under the chromatic model.

The WN in this pulsar is featured in \myfig{fig:ecorr_posteriors} as it has a change in significance at 800 MHz with the GASP backend. This pulsar has newly significant ECORR values supporting the need for a more sophisticated SW model. The RN has a significant decrease in amplitude and a significant increase in spectral index, bringing it more in line with the common process. 

\begin{figure}
    \centering
    \includegraphics[width=0.45\textwidth]{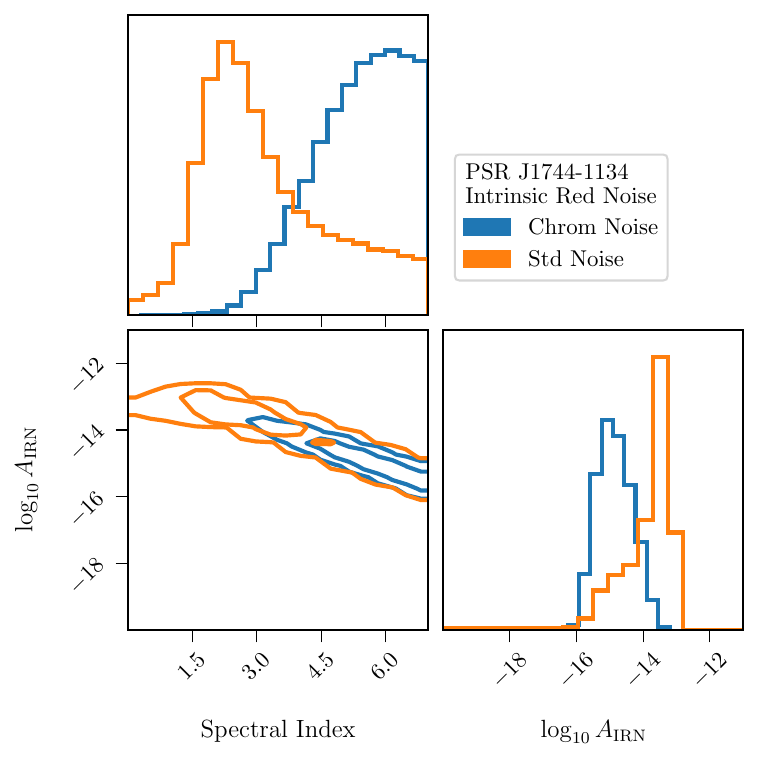}
    \caption{Comparison of RN posteriors for \psrminus{J1744}{1134}. Notice that there is a small amount of support at $\log_{10}A_{\rm RN}\sim-15$ in the standard analysis, that then becomes more significant when the TD chromatic models are used.}
    \label{fig:J1744}
\end{figure}

\subsubsection{\psrplus{B1855}{09}({J1857}$+${0943})}
The chromatic model for \psrplus{B1855}{09} included a significant DMGP with a SE kernel and a significant Chrom GP with a QP kernel. 

The WN in this pulsar has no significant change, and the RN remains the same.

\subsubsection{\psrminus{J1909}{3744}}
The chromatic model for \psrminus{J1909}{3744} has a significant DMGP with a QP kernel, a significant Chrom GP with a QP kernel. This pulsar also has an ecliptic latitude $<30^\circ$, which means that this pulsar may need a more sophisticated SW model than the one presented in this paper. This pulsar is studied further in \citet{Larsen+2024}, where it was found that applying a custom Chrom GP had very little changes to the achromatic RN. 

The WN in this pulsar has no significant change at 800 MHz with the GUPPI backend. The RN has no significant change. 

\subsubsection{\psrplus{J1910}{1256}}
The chromatic model for \psrplus{J1910}{1256} has a significant DMGP with a SE kernel, and no significant Chrom GP. 

The WN in this pulsar is featured in \myfig{fig:ecorr_posteriors} as it has a change in significance with the L-wide PUPPI backend. The RN has no significant change.

\subsubsection{\psrplus{B1937}{21}({J1939}$+${2134})}
\label{sec:B1937}
\begin{figure}
    \centering
    \includegraphics[width=0.45\textwidth]{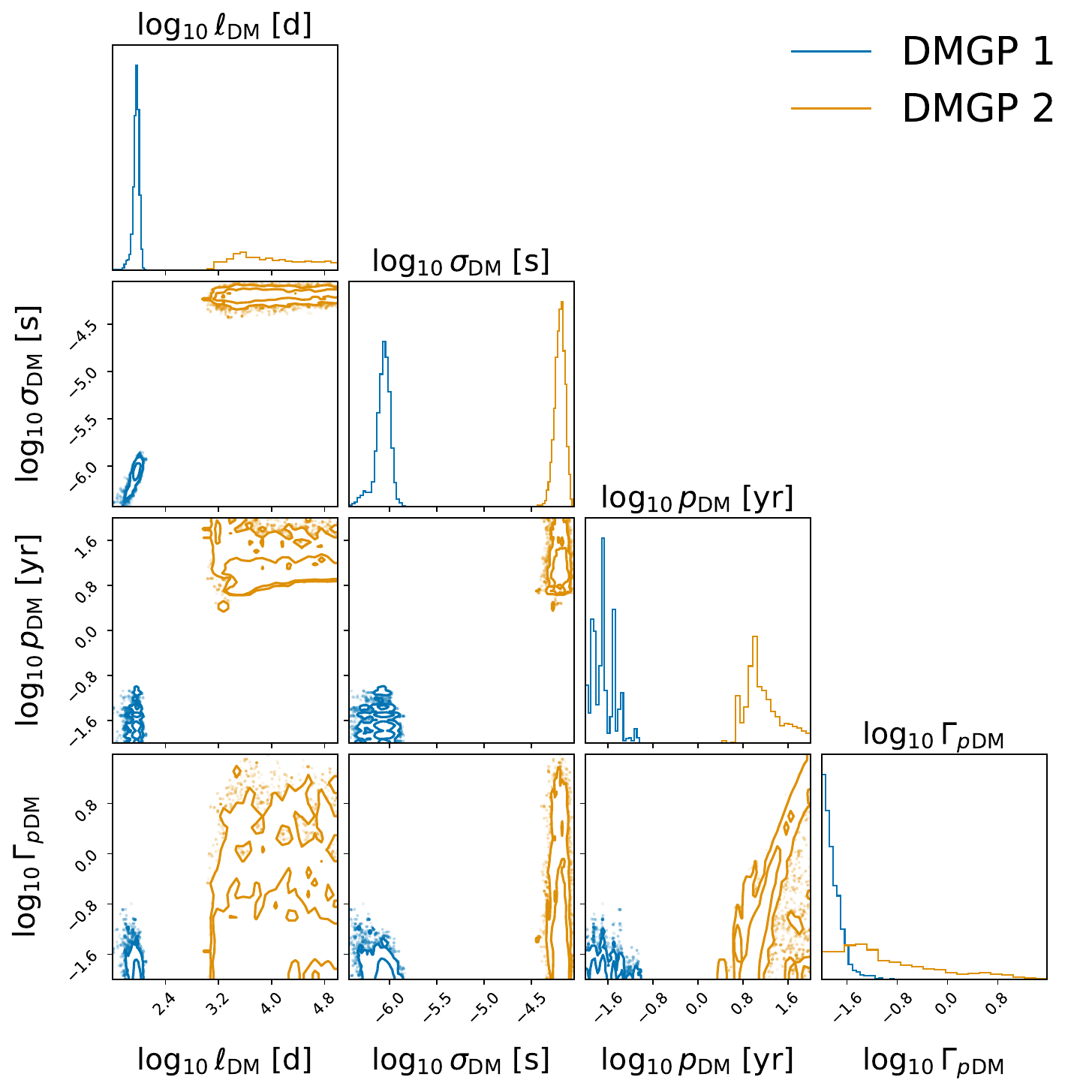}
    \caption{Posteriors distributions for quasi-periodic DM kernel parameters for \psrplus{B1937}{21}. This pulsar represents a unique case where multimodality was recovered in the quasi-periodic DMGP parameters, meriting the use of two independent GPs with tailored prior distributions to fully model all DM variations. The small values of $\log_{10}\ell_{\text{DM}}$ and $\log_{10}p_{\text{DM}}$, corresponding to DMGP 1 (blue), indicate a short timescale DM process. This is simultaneous with a much longer timescale DM process corresponding to DMGP 2 (orange).}
    \label{fig:B1937}
\end{figure}

The chromatic model for \psrplus{B1937}{21} includes \emph{two independent} significant DMGP models, both with a QP kernel, a significant Chrom GP with a SE kernel, and the SW model. The DMGP posteriors for \psrplus{B1937}{21} are shown in \myfig{fig:B1937}. The multimodality seen in the quasi-periodic DMGP parameters lead to the inclusion of two independent DMGP models. Note in particular the differences in the timescales, $\ell_{\rm DM}$,  amplitudes, $\sigma_{\rm DM}$, and periods, $p_{\rm DM}$. This pulsar required a finer basis size $dt=3$ compared to the majority of pulsars that had $dt=15$. This is unsurprising since this pulsar is observed by multiple telescopes and has many observing epochs within a few days of each other.

The WN in this pulsar is featured in \myfig{fig:ecorr_posteriors} as it has a change in significance with the L-wide ASP and the 1.2 GHz GASP backends. The L-wide PUPPI, 1.2 GHz GUPPI, and S-wide ASP have significant decreases ($> 1\sigma $). This pulsar appears in \myfig{fig:WNcompare} and is mentioned as having significant reductions in EFAC, EQUAD, and ECORR.  The remaining backends have no significant change. The RN has no significant change. 

\subsubsection{\psrminus{J2010}{1323}}
The chromatic model for \psrminus{J2010}{1323} has a significant DMGP with a QP kernel, and no significant Chrom GP. This pulsar also has an ecliptic latitude $<30^\circ$, which means that this pulsar may need a more sophisticated SW model than the one presented in this paper.

The WN has no significant changes. The RN has no significant change.

\subsubsection{\psrplus{J2043}{1711}}
The chromatic model for \psrplus{J2043}{1711} has a significant DMGP with a QP kernel and no significant Chrom GP. This pulsar appears in \mytab{tab:det_params} due to having a significant for deterministic DM cusp ($\nu^{-2}$), and scattering cusp ($\nu^{-4}$). 

The WN in this pulsar is featured in \myfig{fig:ecorr_posteriors} as it has a change in significance at 430 MHz with the PUPPI backend and with the L-wide PUPPI backend. The RN in this pulsar is highlighted in \myfig{fig:RNcompare} as the significant red noise from the standard noise model had a change in significance with the custom noise model, making it insignificant.

\psrplus{J2043}{1711} was studied more closely in \myfig{fig:j2043_ecorr} where we investigated the basis size with $dt=1$, $dt=3$, and $dt=15$ with the standard DMX model for the two backends that were shown to have significant ECORR increases in \myfig{fig:ecorr_posteriors}. The two backends were shown to become insignificant again when we used a smaller basis size, leading to an understanding that a finer resolution in the model may mitigate the rise in ECORR seen with Chrom GP.

This pulsar was also studied very recently in \citet{Donlon_2025_J2043} as it had a peculiar acceleration that could be due to a long-period orbital companion or a stellar flyby.

\subsubsection{\psrplus{J2317}{1439}}
The chromatic model for \psrplus{J2317}{1439} has a significant DMGP with a QP kernel, a significant Chrom GP with a SE kernel. This pulsar appears in \mytab{tab:det_params} due to having significance for deterministic signals from Cusp (scattering) DM ($\nu^{-2}$). 

The WN in this pulsar is featured in \myfig{fig:ecorr_posteriors} as it has a change in significance at 430 MHz with the PUPPI backend. This pulsar has newly significant ECORR values supporting the need for a more sophisticated SW model. The RN has no significant change. 

\section{Impacts on Common Process Spectral Characterization}

The common uncorrelated process (CURN) detected with high significance in the NANOGrav $12.5$-year data set (NG12.5), \cite{arz+2020gwb}, was subsequently found to show evidence for being a gravitational wave background in the NANOGrav 15-year data set (NG15), \citep{ng15gwb,ng15data} and other PTA data sets from around the world \citep{eptadr2_3:gwb,pptadr3:gwb, mpta4.5:gwb}. In \cite{ng15gwb}, it was shown that using a Fourier-basis DMGP + other minimal chromatic models (the chromatic ``dips'' in \psrplus{J1713}{0747}) slightly changes the spectral recovery of the GWB, moving the amplitude slightly lower and the spectral index to a slightly steeper values. NANOGrav is completing a full chromatic modeling project on NG15 (Agazie, et al. in prep.), but here we show the consequences of changing out the 20 NG12.5 pulsars in this work on a CURN/GWB recovery. The same 47 pulsars were used in these analyses as in \citet{arz+2020gwb}, swapping out the standard noise models for the chromatic in the subset of 20 pulsars treated in this work. In order to minimize the computational resources needed to do the analysis, a fixed-point analysis was carried out (similarly to \citealt{ng15singlesource}), setting the value of the chromatic models to their maximum a posteriori values. Both a varied spectral index analysis and a free spectral analysis were carried out on this version of the model. The varied spectral index analysis was carried out twice, using the 5 and 30 lowest frequencies, starting with 1/timespan of the data set. 

\begin{figure}
    \centering
    \includegraphics[width=0.45\textwidth]{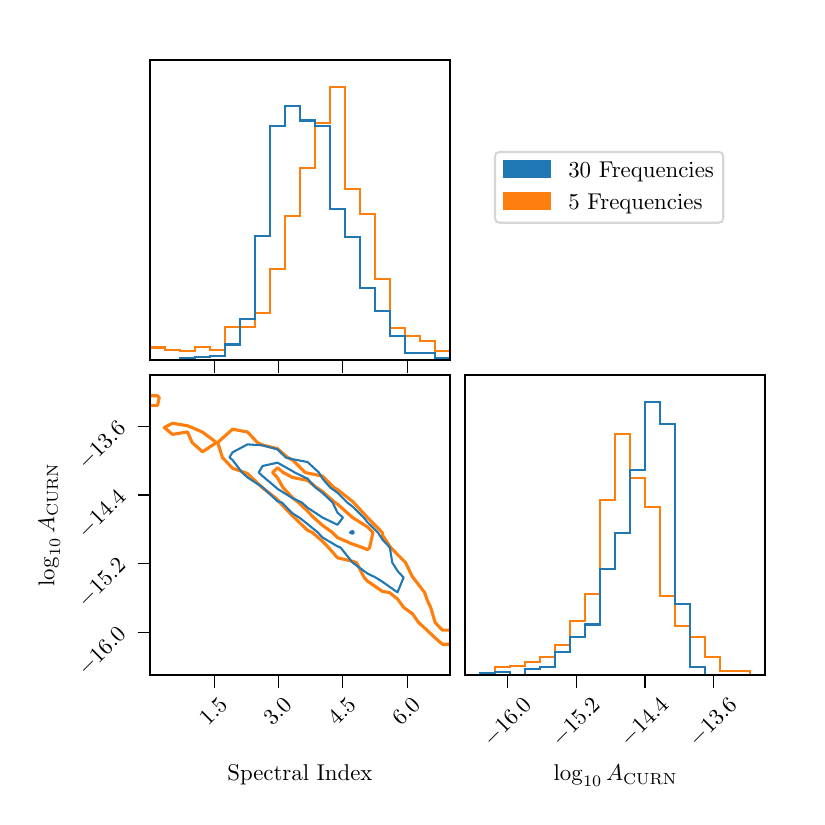}
    \caption{Corner plots comparing iterations using the 5 (orange) and 30 (blue) lowest frequencies to search for a common uncorrelated red noise (CURN) process using the chromatic models with varied amplitude and spectral index. }
    \label{fig:var_gamma_corner1}
\end{figure}

\begin{figure}
    \centering
    \includegraphics[width=0.45\textwidth]{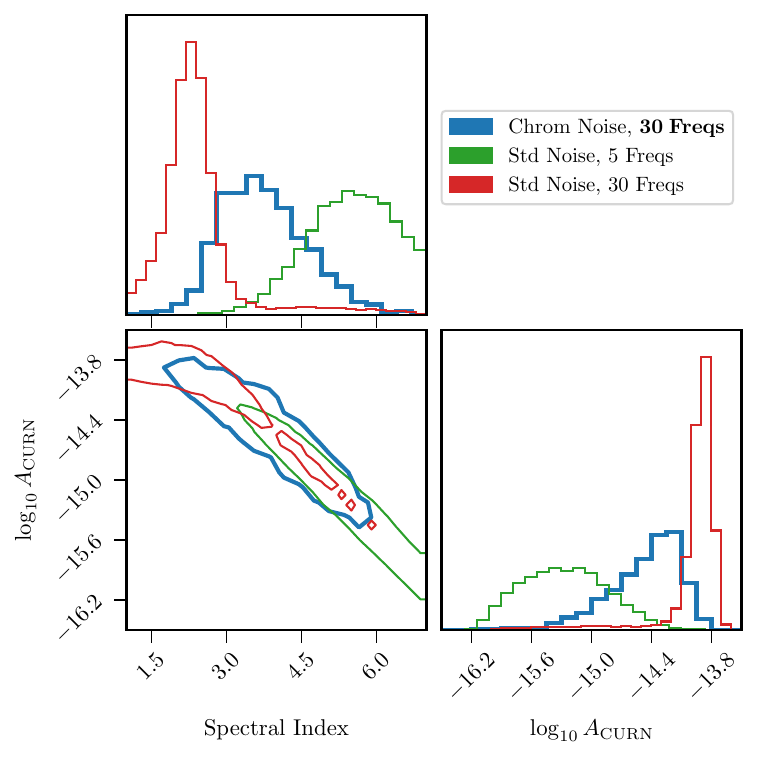}
    \caption{Corner plots comparing various iterations of a search for a common uncorrelated red noise (CURN) process with varied amplitude and spectral index. Specifically, we compare an analysis using the chromatic models with \textbf{30 frequencies} (blue) to the two analyses from \citet{arz+2020gwb} using 5 (green) and 30 frequencies (red).}
    \label{fig:var_gamma_corner2}
\end{figure}

\begin{figure}
    \centering
    \includegraphics[width=0.45\textwidth]{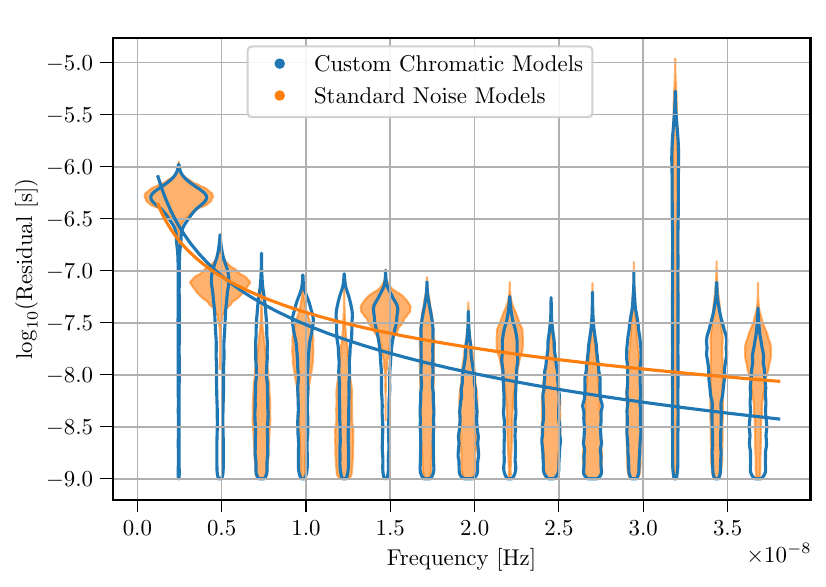}
    \caption{Free spectral analysis of a CURN comparing the standard NG12.5 year analysis to the custom chromatic noise models.  }
    \label{fig:curn_freespec}
\end{figure}

The most important result of this full-PTA analysis is shown in \myfig{fig:var_gamma_corner1}. Recall that in \citet{arz+2020gwb}, the number of frequencies used in the CURN/GWB searches was limited to the five lowest frequencies because unmitigated noise effects were biasing the recovery towards higher amplitude and shallower spectral indices, see \myfig{fig:var_gamma_corner2}. 
As can be seen in \myfig{fig:var_gamma_corner1}, the recovery of the amplitude and spectral index with the chromatic models is not as biased by the number of frequencies as the standard noise models. The 5-frequency search returns broader posteriors, which should be expected because there should be more information in 30 frequencies, but the posteriors, especially the joint posterior, are largely overlapping. Compare this to the varied spectral index results in \myfig{fig:var_gamma_corner2} from the standard model used in \citet{arz+2020gwb}, where the choice of frequencies strongly biases the recovery of the parameters.
The results from the free spectral analysis, shown in \myfig{fig:curn_freespec} echo the same results. The width of the violins represents the probability of the recovered posteriors at each frequency. The figure reveals that the recovery of power in the 2nd, 6th, 9th and 15th frequency bins is less significant in the model using the 20 new chromatic models. These support a steeper common process across the frequency range. 
Searches for Hellings-Downs (HD) spatial correlations between the pulsars were also carried out \citep{hourihane+2023}. We resampled the chains from the CURN run using a likelihood model that included HD correlations and also ran an optimal statistics analysis \citep{anholm+2009os,chamberlin+2015os,vite18}. In both cases, no substantial support for spatial correlations was found in the data sets. Seeing lower support for CURN power at all frequencies except the 1st frequency bin in \myfig{fig:curn_freespec} could point to this lack of support. It is important to point out again that less than half of the pulsars went through the full chromatic model selection process in this analysis. The forthcoming reanalysis of NG15 customizing noise models for all pulsars may see more dramatic effects on correlation recovery.

\section{Discussion}

The complex interplay between white noise, specifically ECORR or per-epoch WN, chromatic noise, which rises from the turbulent ISM, and achromatic red noise, which includes the GWB, is well documented and continues to be an area of robust scientific endeavor. The work presented in this paper represents the culmination of years of effort, and while the results cannot be distilled down into a single, golden rule for pulsar timing noise analyses, there are a handful of important takeaways and intriguing findings, which point toward the next series of steps to be undertaken in this line of work. We will delve into these throughout the rest of this section. However, the most important thing to remember is that these findings are based on what is now an archival data set, and the most impactful insights will come from applying these models and lessons learned to newer and more sensitive data. 

\subsection{Chromatic Models \& ECORR \label{sec:chrom_models_ecorr}}

Looking at \mytab{tab:ecorr_params} and \myfig{fig:ecorr_posteriors} one can see that the new chromatic models have an effect on many of the ECORR WN parameters. ECORR tracks changes in the TOAs correlated on short, intra-epoch timescales. Any timing effect that stochastically changes the arrival time of pulses can be modeled by ECORR, if it is \emph{not} accounted for by other parts of the timing model. Any remaining radio frequency-dependent noise, due to mis-modeled DM variations or unmodeled scattering variations, will be correlated across the TOAs in a given receiver. It is evident in this work and other work done on NANOGrav data where DMX is removed \citep{Larsen+2024} that the immense freedom of the DMX model allows many chromatic effects to be mitigated, including DM/SW variations, but also scattering and other chromatic events, like the purported profile change of \psrplus{J1713}{0747}\citep{lam_j1713_2nd_2018, goncharov+2020, atel14642_J1713dip, atel14652_J1713dip, lam_j1713_3rd_2021, lin_j1713_3rd_2021, singha_j1713_3rd_2021, jennings_j1713_3rd_2024}. In this investigation, we observed that when DMX is removed and only a DMGP is used to model the noise in pulsars, ECORR often increases.
For the final custom noise models selected for these 20 pulsars there are three scenarios for the ECORR values:
\begin{enumerate}
	\item The ECORR values do not change significantly between the DMX model and the CNM. The vast majority of ECORR values fall into this category.
	\item The ECORR values decrease by $>1\sigma$ when employing the CNM or become insignificant. In all cases where this happens, except one, the data prefer a scattering model. This points to the cause of at least some ECORR-modeled noise in the standard noise model being from previously unmodeled scattering variations.
	\item The ECORR values increase by $>1\sigma$ when employing the CNM or become significant. This is less than ideal, but can be explained in a few cases (see below) and those explanations point the way for pipeline development for future analyses. However, further mitigation of ECORR is beyond the scope of this investigation on the now outdated NG12.5.
\end{enumerate}

Now let us turn to possible reasons for the cases where the ECORR values increase significantly. The most obvious is that in some cases a more finely sampled model is needed. The $dt$ parameter sets the size of the linear interpolation basis used for the time-domain models and, as shown in \mysec{sec:basis_size}, we are free to decrease $dt$ at the cost of using a larger basis size. In order to limit the scope of the current project originally all pulsars used $dt=15$ days except \psrplus{B1937}{21} and \psrplus{J1713}{0747}. Once well established models were found for these two pulsars, model selection was done between $dt=[3,7,15]$ days. In both cases the smallest, $dt=3$ days, was preferred. While these pulsars have the largest data sets it was deemed worthwhile to use such large basis sizes since they are observed by both the GBT and AO and often have cross telescope epochs within a few days of each other. In both cases a few of the ECORR parameters decreased as $dt$ decreased. 

\subsubsection{Time-Domain Kernel Basis Size}\label{sec:basis_size}

\begin{figure}
    \centering
    \includegraphics[width=0.45\textwidth]{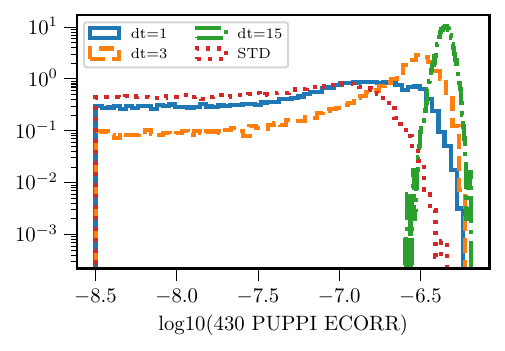}
    
    \includegraphics[width=0.45\textwidth]{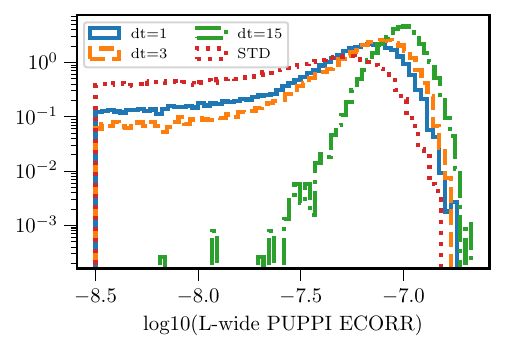}
    \caption{\psrplus{J2043}{1711} ECORR posterior comparison. Posteriors for two different receiver-backend combinations are shown for three choices of $dt$ and compared to the result using the standard DMX model. }
    \label{fig:j2043_ecorr}
\end{figure}

In order to investigate that too-large interpolation bins could be responsible for higher ECORR values we analyzed \psrminus{J1024}{0719}, \psrplus{J1741}{1351} and \psrplus{J2043}{1711} with $dt=1$ and $dt=3$. For the first two pulsars none of the ECORR values changed appreciably with basis size. However, as seen in \myfig{fig:j2043_ecorr}, the two ECORR values that increased for \psrplus{J2043}{1711} again became insignificant when we used a smaller basis size. This demonstrates that, at least in some cases, a finer resolution to the model basis will mitigate the rise in ECORR seen with GP chromatic model. 

\subsubsection{Impacts of a Rigid SW Model}

Another possible culprit to increasing the ECORR in these custom chromatic noise models is a lack of freedom where the SW modeling is concerned. As shown in \citet{hazboun+2022sw, tiburzi+2021, susarla+2024} pulsars that are close to the ecliptic plane may see more variations in the density of solar electrons, or even solar events where large quantities of electrons are jettisoned from the Sun. This work and the fact that a number of the pulsars with newly significant ECORR values (\psrplus{J0030}{0451},  \psrminus{J1744}{1134}, \psrplus{J2317}{1439}) and increasing ECORR values, see \mytab{tab:ecorr_params}, (\psrplus{J0645}{5158}, \psrminus{J1614}{2230}, \psrplus{J1713}{0747}, \psrplus{J1738}{0333}, \psrminus{J1909}{3744}, \psrminus{J2010}{1323}) have ecliptic latitudes $<30^\circ$ supports the possibility that these pulsars need a more sophisticated SW model than the binned-piecewise one used here. Efforts are currently focused on adding more sophisticated SW models, like those in \citet{nitu+2024, susarla+2024}, for NG15.

\subsection{Chromatic Models \& Red Noise}

In regards to searches for gravitational waves, particularly the stochastic gravitational wave background, one of the most important effects of the chromatic models that we see on the pulsars are the changes in the RN parameter recovery. Figures~\ref{fig:RNcompare} and \ref{fig:var_gamma_corner2} highlight the changes that the new models make in the individual pulsar analyses and the full search for a CURN process, respectively. The net result of the changes in the RN posteriors supports a steeper RN process in the full CURN analysis, which is possibly brought about by one or more strongly overlapping effects, including:

\begin{itemize}
    \item Decreasing the (white) noise floor across the spectrum. See for example \psrplus{J1713}{0747}, where the chromatic dips caused by profile events add a substantial amount of white noise into the data.
    \item Absorbing mismodeled chromatic red noise into ``physically appropriate'' channels. This is certainly the case for the pulsars where scattering models are preferred, but can also be seen in SW models or subtle changes in other noise process parameters.
    \item Mitigating high frequency noise to allow for steeper RN recovery, either by absorbing noise in the WN sector of the model or using sufficiently high frequency chromatic models to mitigate chromatic noise. See for example \psrminus{J0613}{0200} where the decrease in ECORR and the addition of a number of chromatic model components, is paired with an increase in the spectral index. 
\end{itemize}

The movement of noise power between the different components of the model is well documented \citep[e.g.,][]{hazboun:2020slice,Larsen+2024,inpta_dr1_noise} and the use of physically motivated chromatic models helps to keep the noise out of the more generic RN power law channel.

\subsection{Model Selection for Pulsar Noise}

Chromatic models for pulsar timing data date back to the earliest observations of their radio signals \citep{bell+1968,counselman+1970}. In the past, DM models (including long-timescale variations, SW variations and annual variations) have sufficed, but higher precision data, and the concerted campaigns of PTAs, have increased the need for models that take into account smaller and smaller chromatic effects. The numerous recent noise-model analyses of PTA data sets each tackle the model selection question in various ways \citep{inpta_dr1_noise,eptadr2_2:noise,pptadr3:noise,Miles+2025_noise}, usually employing Bayesian model selection techniques to compare various models against each other. All such analyses need to take into account the conflict between the Bayesian aspiration to ``model it all and let the data decide'' and the enormous computational costs of model selection \citep{vH2025_model_averaging}. There has been recent advances in both the methods used for PTA data set models (e.g., ``spike-slab'' priors to mimic a transdimensional model, \citep{vH2025_model_averaging}) and the increasing availability of GPUs and the software to interface with them, however, this analysis predates these new developments. 

Here we list highlights from the current scheme for model selection that works towards an efficient method, given the tools available.

1. The construction of the time-domain kernels as a nested model makes testing for added complexity fairly simple. Simply by looking at the posterior recovery of the various amplitude parameters we get a broad idea which parts of the model are significant. Fourier domain GPs can also be tested in this same way. However, testing between TD and FD models will require some sort of model selection.  
    
2. The interplay between CNMs and the SW model is subtle and requires iterative analyses. The SW model is still important for off-ecliptic pulsars, and therefore should be included as a global parameter (or global plus tailored parameter) in all individual pulsar noise runs.
    
3. The interplay of ECORR and CNMs is very evident from this work. As we layer on higher order (e.g., scattering) CNMs, use better SW models and more finely sample the CNMs (e.g., linear interpolation bases with smaller $dt$s) we see that ECORR values decrease. In the forthcoming analysis of the NG15 data set more custom modeling (SW GPs, varied index for higher order CNMs, etc.) leads to even better amelioration of ECORR. 

\subsection{Next Steps}
A full chromatic noise-model analysis, that includes both these time-domain models and the more traditional frequency-domain based GPs is currently underway on the NG15 data set. Many of the techniques used herein are extended to use Bayesian techniques to select between the two classes of models as well as the resolution of the analysis (whether $dt$ or the number of frequencies), frequency-dependent ``band noise'' and the chromaticity (index of radio frequency dependence) for higher order chromatic terms, just as has been done by other PTAs in recent years. These types of full classifications will help immensely with efficiently carrying out noise analyses for the forthcoming IPTA DR3 data set.

The stationary time-domain kernels, $k(\tau)$, used herein are related to structure functions, $D(\tau)=2[k(0)-k(\tau)]$, \citep{lcc+15, jones+2017_ng9_dm}, and thus the posteriors represent a physical description of the ISM processes/timescales. One power of the TD kernels then is our ability to place meaningful priors on them from other ISM studies/observations. This would be an incredibly interesting line of research to take forward. Right now, few other DM models (outside of the SW) have any kind of physically-informed priors. Further study is required to figure out how to effectively do this. Recent work \citep{geiger+2025} has shown that the chromaticity of scattering delays can change epoch to epoch, which would make modeling these higher chromatic terms particularly challenging. Thankfully, the upcoming implementation of cyclic-spectroscopy backends \citep{turner+2025} will allow us to measure the chromaticity and may even mitigate the majority of these higher-order chromatic effects.

\begin{contribution}
All authors contributed to the activities of the NANOGrav
collaboration leading to the work presented here and reviewed the manuscript, text, and figures prior to the paper’s submission. Additional specific contributions to this paper are as follows. J.S.H and J.S developed the project, organized the analyses and the writing of the manuscript. J.B., J.S.H., B.L., D.J.O., and J.S. wrote the manuscript, developed the figures and carried out aggregate analyses on pulsar results. P.B., J.A.E., J.S.H., J.S., and S.R.T. developed the models and analysis software for this work. J.B., B.B., S.C., J.S.H, A.M.H., K.I., A.J., A.R.K., N.L., B.L., N.S.P., M.Y.K., M.S., B.J.S., J.P.S., J.S., C.A.W., J.V., and C.Y. ran analyses on individual pulsars. J.B., P.T.B., S.C., J.M.C., R.J., J.K., J.S.H., M.T.L., B.L., D.R.M., C.M.F.M., D.J.O., X.S., J.S., D.R.S., S.R.T., J.T., and M.V. worked to develop and interpret the analyses in the larger context of pulsar astrophysics. A.D.H. derived the power spectral density of the quasiperiodic kernel. Z.A., H.B., P.R.B., H.T.C., M.E.D., P.B.D., T.D., J.A.E., R.D.F., E.C.F., E.F., N.G., P.A.G., D.C.G., M.L.J., M.T.L., D.R.L., J.L., R.S.L., M.A.M., C.N., D.J.N., T.T.P., N.S.P., S.M.R., R.S., I.H.S., K.S., J.K.S. and S.J.V. developed the $12.5$-year data set through a combination of observations, arrival time calculations,
data checks and refinements, and timing-model development and analysis.
\end{contribution}

\begin{acknowledgments}
\input{acks12p5yrCNM} 
\end{acknowledgments}

\facilities{Arecibo, GBT, VLA}

\software{
    astropy \citep{astropy},
    matplotlib \citep{matplotlib},
    numpy \citep{numpy},
    enterprise \citep{enterprise},
    enterprise\_extensions \citep{enterprise_ext},
    ptmcmcsampler \citep{ptmcmcsampler},
}

\bibliographystyle{aasjournalv7}
\bibliography{hazgrav}

\appendix
\section{Power spectral Density of the Quasi-Periodic Kernel}
\label{appendix:QP_kernel_PSD}

The quasi-periodic kernel in  \myeq{eq:qp_kernel} can be split into a squared exponential component and a global periodic component,
\begin{eqnarray}
    K(t) &=P(t)L(t)\;,
\end{eqnarray}
each of which we treat separately, where $P(t) =  e^{-\Gamma \sin^2(\frac{\pi}{p}t)}$ and $L(t) = \sigma^2 e^{-\frac{t^2}{2 \ell^2}}$. We remove the quadratic sine via a double angle replacement, and replace period with angular frequency, $\omega_0 = \frac{2\pi}{p}$, for neater notation giving $P(t) = e^{-\frac{\Gamma}{2}} e^{\frac{\Gamma}{2}\cos(\omega_0 t)}$. A hyperbolic trigonometric substitution,
\begin{equation*}
    e^{-\frac{\Gamma}{2}} \left[e^{\frac{\Gamma}{2}\cos\omega_0 t}\right] = e^{-\frac{\Gamma}{2}} \left[\cosh\left(\frac{\Gamma}{2}\cos\omega_0 t\right) + \sinh\left(\frac{\Gamma}{2}\cos\omega_0 t\right) \right]\;,
\end{equation*}

and the following series representations,

\[
\cosh x 
= \sum_{n}^{\infty} \frac{1}{2n} x^{2n}
\quad,\quad
\sinh x 
= \sum_{n}^{\infty} \frac{1}{2n+1} x^{2n+1}\;,
\]
Allow us to write the expression as
\begin{eqnarray}
    P(t)
    &= 
    e^{-\frac{\Gamma}{2}}
    \left[
        \sum_{n}^{\infty} \frac{1}{(2n)!} \left(\frac{\Gamma}{2}\right)^{2n} \left(\cos\omega_0 t \right)^{2n}
    \right.\nonumber\\
    &+
    \left. 
    \sum_{n}^{\infty} \frac{1}{(2n+1)!} \left(\frac{\Gamma}{2}\right)^{2n+1}\left(\cos\omega_0 t \right)^{2n+1}
\right]\;.
\end{eqnarray}
Powers of trigonometric functions have a convenient representation, however they are different depending on whether the power is even or odd, 
\[
(\cos\omega_0 t)^{2n+1} = \frac{1}{2^{n}}\sum_{k=0}^{n} \frac{(2n+1)!}{k!(2n+1-k)!} \cos(2\omega_0 t(n-k+\frac{1}{2}))
\]
\[
(\cos\omega_0 t)^{2n} = \frac{1}{2^{2n}}\frac{(2n)!}{n!n!} + \frac{2}{2^{2n}}\sum_{k=0}^{n-1} \frac{(2n)!}{k!(2n-k)!}\cos(2\omega_0 t(n-k))\;.
\]
Inserting these expressions into our expression for $P(t)$ we have 
\begin{eqnarray}
    P(t)
    &=
    e^{-\frac{\Gamma}{2}}
    \left[
        \sum_{n}^{\infty} \frac{1}{(2n)!}   
        \left(
            \frac{\Gamma}{2}
        \right)^{2n}
        \left(
            \frac{1}{2^{2n}}\frac{(2n)!}{n!n!} + \frac{2}{2^{2n}}\sum_{k=0}^{n-1} \frac{(2n)!}{k!(2n-k)!}\cos{(2\omega_0 t(n-k))}
        \right)
    \right.\nonumber\\
    &+ 
    \left. 
        \sum_{n}^{\infty} \frac{1}{(2n+1)!} 
        \left(
            \frac{\Gamma}{2}
        \right)^{2n+1}
        \left(
            \frac{1}{2^{n}}\sum_{k=0}^{n} \frac{(2n+1)!}{k!(2n+1-k)!} \cos{(2\omega_0 t(n-k+\frac{1}{2}))}
        \right)
    \right]\;.
\end{eqnarray}
Simplifying and replacing the variable \(\frac{\Gamma}{2} = A\) we have 
\begin{eqnarray}
    P(t)
    &=
    &e^{-A}
    \left[
        \sum_{n}^{\infty} \frac{1}{n!n!} \left(\frac{A}{2}\right)^{2n}
        \right. \nonumber \\
        &&+
        \left.
        2\sum_{n=0}^{\infty}\sum_{k=0}^{n-1}
        \left(
            \frac{A}{2} 
        \right)^{2n}
        \frac{1}{k!(2n-k)!} \cos{(2\omega_0 t(n-k))}
        \right. \nonumber \\
        &&+
        \left.
        2\sum_{n=0}^{\infty}\sum_{k=0}^{n}
        \left(
            \frac{A}{2} 
        \right)^{2n+1}
        \frac{1}{k!(2n+1-k)!} \cos{(2\omega_0 t(n-k+\frac{1}{2}))}
    \right]
\end{eqnarray}
We can define a new index \(m = n-k\), which implies \(n=m+k\), throughout our expression. We can also shift over an index in the second term so our second summations match up
\begin{eqnarray}
    P(t)
    &=&e^{-A}
    \left[
        \sum_{m+k=0}^{\infty} \frac{1}{(m+k)!(m+k)!} \left(\frac{A}{2}\right)^{2k+2m}
        \right. 
        \nonumber \\
        &&+
        \left.
        2\sum_{m+k=1}^{\infty}\sum_{k=0}^{m+k}
        \frac{1}{k!(k+2m)!} 
        \left(
            \frac{A}{2} 
        \right)^{2k+2m}
        \cos{(2\omega_0 t m)}
        \right. 
        \nonumber\\
        &&+
        \left.
        2\sum_{m+k=0}^{\infty}\sum_{k=0}^{m+k}
        \frac{1}{k!(k+2m+1)!} 
        \left(
            \frac{A}{2} 
        \right)^{2k+2m+1}
        \cos{(2\omega_0 t(m+\frac{1}{2}))}
    \right]
\end{eqnarray}
The terms in these series match those in the  defining series for the hyperbolic Bessel functions of the First Kind, with \(\alpha\) being either $2m$ or $2m+1$,
\begin{eqnarray}
    I_\alpha(z) =  \sum_{k=0}^{\infty} \frac{1}{k!(k+\alpha)!} 
    \left(
        \frac{z}{2}
    \right)^{2k+\alpha}
\end{eqnarray}
    
Since all our summation indices are themselves a sum we have the freedom to let $k$ be the variable over which we sum over first. This lets us substitute in the Bessel functions directly. The first term has no independent $m$ indicies so we can set $m=0$ with no loss of generality 
\begin{eqnarray}
    P(t)
    =
    e^{-A}
    \left[
        I_0(A)
        +2\sum_{m=1}^{\infty} I_(A){2m} \cos{(2\omega_0 t m)}
        +2\sum_{m=0}^{\infty} I_{2m+1}(A) \cos{(2\omega_0 t(m+\frac{1}{2}))}
    \right]
\end{eqnarray}
In this representation it will be much easier to perform the Fourier transform of our entire kernel. So we multiply by the local "squared exponential" component
\begin{eqnarray}
    K(t)
    &=&     
    I_0(A)\sigma^2 e^{-\frac{\Gamma}{2}} 
    \left[
    e^{-\frac{t^2}{2 \ell^2}}
    \right]
    \nonumber \\
    &&+
    2I_1(A)\sigma^2 e^{-\frac{\Gamma}{2}} 
    \left[
    e^{-\frac{t^2}{2 \ell^2}}\cos{\omega_0 t}
    \right]
    \nonumber \\
    &&+
    2\sum_{m=1}^{\infty} 
    I_{2m}(A) \sigma^2 e^{-\frac{\Gamma}{2}} 
    \left[
        e^{-\frac{t^2}{2 \ell^2}} \cos{(2\omega_0 t m)} 
    \right]
    \nonumber \\
    &&+
    2\sum_{m=1}^{\infty} I_{2m+1}(A) \sigma^2 e^{-\frac{\Gamma}{2}} 
    \left[
    e^{-\frac{t^2}{2 \ell^2}} \cos{(2\omega_0 t(m+\frac{1}{2}))}
    \right]
\end{eqnarray}
Applying the Fourier transform \(\mathcal{F}\) to this representation and remembering that the Hyperbolic Bessel Functions are simply evaluated at a single point and not the full functions, we get
\begin{eqnarray}
    \mathcal{F}[K(t)]
    &=&     
    I_0\left(\frac{\Gamma}{2}\right)\sigma^2 e^{-\frac{\Gamma}{2}} 
    \mathcal{F}\left[
    e^{-\frac{t^2}{2 \ell^2}}
    \right]
    \nonumber \\
    &&+
    2I_1\left(\frac{\Gamma}{2}\right)\sigma^2 e^{-\frac{\Gamma}{2}} 
    \mathcal{F}\left[
    e^{-\frac{t^2}{2 \ell^2}}\cos{\omega_0 t}
    \right]
    \nonumber \\
    &&+
    2\sum_{m=1}^{\infty} 
    \left(
    I_{2m}\left(\frac{\Gamma}{2}\right) \sigma^2 e^{-\frac{\Gamma}{2}} 
    \mathcal{F}\left[
        e^{-\frac{t^2}{2 \ell^2}} \cos{(2\omega_0 t m)} 
    \right]
    \right.
    \nonumber \\
    &&+
    \left.
    I_{2m+1}\left(\frac{\Gamma}{2}\right) \sigma^2 e^{-\frac{\Gamma}{2}} 
    \mathcal{F}\left[
    e^{-\frac{t^2}{2 \ell^2}} \cos{(2\omega_0 t(m+\frac{1}{2}))}
    \right]
    \right)
\end{eqnarray}
Using the unitary Fourier transform defined as \(\mathcal{F}\left[g(t)\right] = \frac{1}{\sqrt{2\pi}}\int_{-\infty}^{\infty} g(t)e^{-i\omega t} \,dt = G(\omega)\), we can find the transform of the local "squared exponential" component of the kernel. The Periodic contribution is similar.

\begin{eqnarray}
    \mathcal{F}
    \left[
        e^{-a t^2}
    \right]
    =
    \frac{1}{\sqrt{2a}} e^{-\frac{\omega^2}{4a}}
    ,\quad
    \mathcal{F}
    \left[
        g(t)\cos{(b)}
    \right]
    =
    \frac{1}{2} G(\omega-b) + \frac{1}{2} G(\omega+b)
\end{eqnarray}

Combining both transforms is straightforward and we end up with a term of the following form

\begin{eqnarray}
    \mathcal{F}
    \left[
        e^{-a t^2}\cos{(b)}
    \right]
    =
    \frac{1}{\sqrt{2a}} e^{-\frac{1}{4a}(\omega^2+b^2)}\cos{
    \left(
        \frac{b}{2a}\omega
    \right)}
\end{eqnarray}

Substituting \(a=\frac{1}{2\ell^2}\) and \(b=0, \ \omega_0, \ 2m\omega_0,  \ (2m+1)\omega_0\)

Applying these to our expression of the Quasi-Periodic Kernel

\begin{eqnarray}
    \mathcal{F}
    \left[
        K(t)
    \right]
    =&
    \ell \sigma^2 e^{-\frac{\Gamma}{2}}&
    \left(
        I_0
        \left(
            \frac{\Gamma}{2}
        \right)
        e^{-\frac{1}{2} \ell^2 \omega^2}
    \right.
    \nonumber\\
    &&+ 
    \left.
        I_1
        \left(
            \frac{\Gamma}{2}
        \right)
        e^{-\frac{1}{2} \ell^2 (\omega^2 + \omega_0^2)} \cos{(\ell^2 \omega_0 \omega)}
    \right.
    \nonumber\\
    &&+
    \left.
    2\sum_{m=1}^{\infty} 
    \left[
    I_{2m}\left(\frac{\Gamma}{2}\right)
    e^{-\frac{1}{2} \ell^2 (\omega^2 + 4m^2 \omega_0^2)} \cos{(2m \ell^2 \omega_0 \omega)}
    \right.
    \right.
    \nonumber\\
    &&+
    \left.
    \left.
    I_{2m+1}
    \left(
        \frac{\Gamma}{2}
    \right)
    e^{-\frac{1}{2} \ell^2 (\omega^2 + (2m+1)^2 \omega_0^2)} \cos{((2m+1) \ell^2 \omega_0 \omega)}
    \right]
    \right)
\end{eqnarray}

Replacing \(\omega_0 = \frac{2\pi}{p}\), we have the full analytic series representation of the Fourier Transform of the Quasi-Periodic Kernel

\begin{eqnarray}
    \mathcal{F}
    \left[
        K(t)
    \right]
    =&
    \ell \sigma^2 e^{-\frac{\Gamma}{2}}&
    \left(
        I_0
        \left(
            \frac{\Gamma}{2}
        \right)
        e^{-\frac{1}{2} \ell^2 \omega^2}
    \right.
    \nonumber\\
    &&+ 
    \left.
        I_1
        \left(
            \frac{\Gamma}{2}
        \right)
        e^{-\frac{1}{2} \ell^2 (\omega^2 + 
        \left(
            \frac{2\pi}{p})^2
        \right)} 
        \cos{(\ell^2 
        \left(
            \frac{2\pi}{p}
        \right) 
        \omega)}
    \right.
    \nonumber \\
    &&+
    \left.
    2\sum_{m=1}^{\infty} 
    \left[
    I_{2m}
    \left(
        \frac{\Gamma}{2}
    \right)
    e^{-\frac{1}{2} \ell^2 (\omega^2 + 4m^2 
    \left(
        \frac{2\pi}{p})^2
    \right)} 
    \cos{(2m \ell^2 \left(
        \frac{2\pi}{p}
    \right)
    \omega)}
    \right.
    \right.
    \nonumber\\
    &&+
    \left.
    \left.
    I_{2m+1}
    \left(
        \frac{\Gamma}{2}
    \right)
    e^{-\frac{1}{2} \ell^2 (\omega^2 + (2m+1)^2 \left(\frac{2\pi}{p})^2\right)} \cos{((2m+1) \ell^2 \left(\frac{2\pi}{p}\right) \omega)}
    \right]
    \right)
\end{eqnarray}

\begin{figure}
    \centering
    \includegraphics[width=0.9\textwidth]{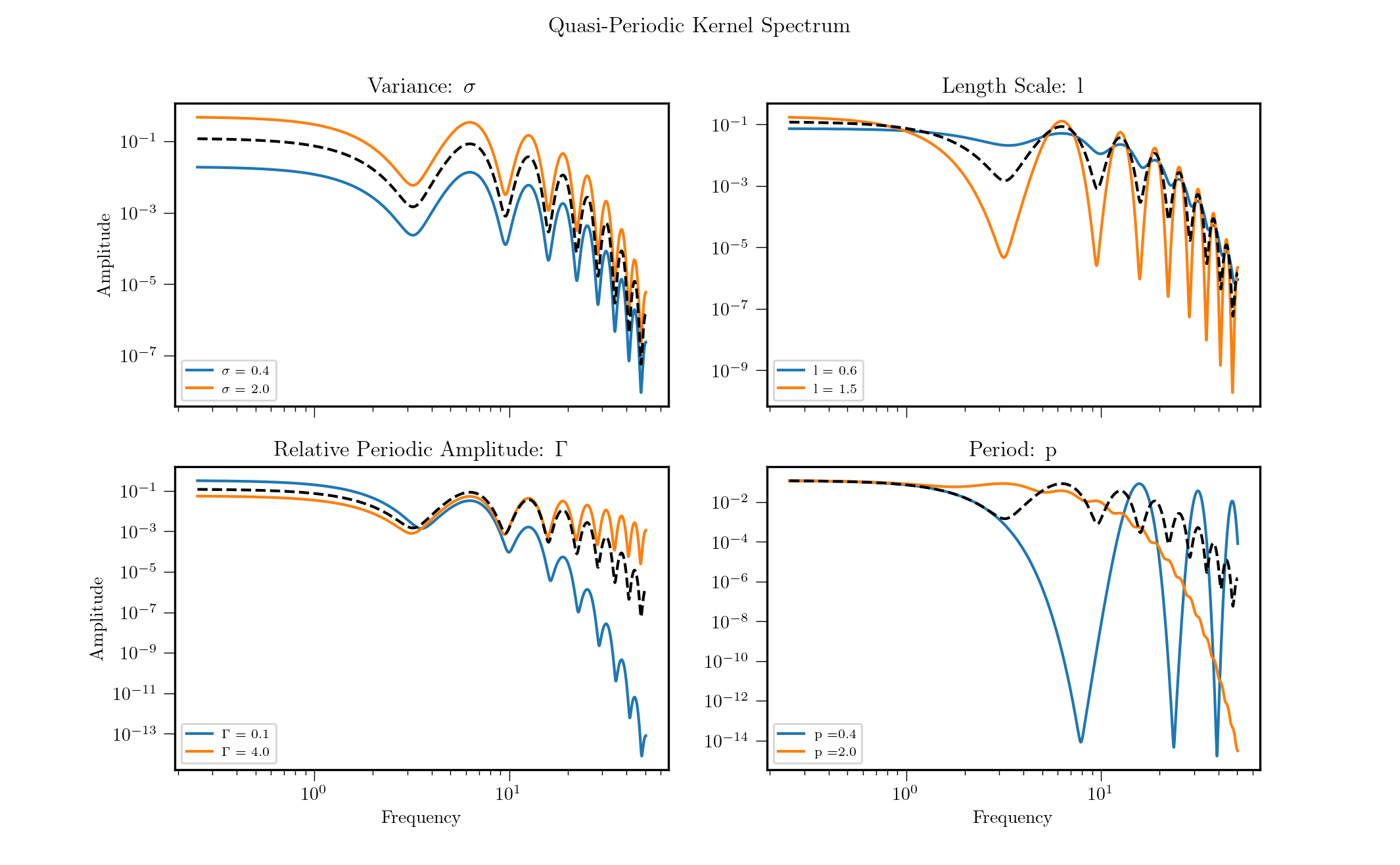}
    \caption{Power spectral density of quasiperiodic time-domain kernel.}
    \label{fig:qp_psd}
\end{figure}

\section{Detailed Analysis Priors \& Results}\label{appendix:tables}

\begin{table}
    \centering
    \begin{tabular}{ c | c c c @{}}
        \hline Model & Parameter & Prior & Units \\
        \hline & EFAC & $\mathcal{U}(0.01, 10.0)$ & \\
        WN & EQUAD & $\log_{10}\mathcal{U}(10^{-8.5}, 10^{-5})$ & sec \\
        & ECORR & $\log_{10}\mathcal{U}(10^{-8.5}, 10^{-5})$ & sec \\
        \hline \multirow{2}{*}{RN} & $A$ & $\log_{10}\mathcal{U}(10^{-20}, 10^{-11})$ & \\
        & $\gamma$ & $\mathcal{U}(0, 7)$ & \\
        \hline & $\sigma$ & $\log_{10}\mathcal{U}(10^{-10}, 10^{-4})$ & sec \\
        & $\ell$ & $\log_{10}\mathcal{U}(10^1,10^4)$ & day \\
        DM \& & $\Gamma_p$ & $\log_{10}\mathcal{U}(10^{-3},10^2)$ & \\
        Chromatic & $p$ & $\log_{10}\mathcal{U}(10^{-2},10^1)$ & yr \\
        & $\alpha_{\mathrm{wgt}}$ & $\log_{10}\mathcal{U}(10^{-4},10^1)$ & \\
        & $\ell_2$ & $\log_{10}\mathcal{U}(10^2,10^7)$ & MHz \\
        \hline \multirow{2}{*}{AV} & $A$ & $\log_{10}\mathcal{U}(10^{-10},10^{-2})$ & sec \\
        & $\phi$ & $\mathcal{U}(0,2\pi)$ & \\
        \hline & $A$ & $\log_{10}\mathcal{U}(10^{-10},10^{-2})$ & sec \\
        Cusp & $\tau_{\mathrm{pre/post}}$ & $\log_{10}\mathcal{U}(10^0,10^{2.5})$ & day \\
        & $t_0$ & $\mathcal{U}(t_{\mathrm{min}},t_{\mathrm{max}})$ & MJD \\
        \hline \multirow{2}{*}{SW} & $n_{E,i}$ & $\mathcal{U}(0,30)$ & cm$^{-3}$ \\
        & $n_E^{(4.39)}$ & $\log_{10}\mathcal{U}(-4,-2)$ & cm$^{-3}$ \\
        \hline
    \end{tabular}
    \caption{Priors and units on noise model parameters during MCMC analyses. The bounds, $t_{\mathrm{min}}$ and $t_{\mathrm{max}}$, on the $t_0$ cusp parameter are determined on a per-pulsar level. Additional tailored priors are used in the models for PSRs \pplus{B1937}{21} and \pplus{J1713}{0747}, as detailed in sections \mysec{sec:B1937} and \mysec{sec:J1713} respectively.}
    \label{tab:priors}
\end{table}

\input{Table_GP_Params_alt2}

\input{Table_DetParams}

\input{Table_RN_Params}

\input{Table_ECORR_Params}

\end{document}

%% file: Table_ModelResults.tex
\begin{table*}[t]
\centering
\begin{tabular}{l|ccccc}
\hline
Pulsar               & T (years) & dt (days) & DM ($\nu^{-2}$) Kernel & Chromatic ($\nu^{-4}$) Kernel & Deterministic Signals                   \\ \hline 
\pplus{B1855}{09}    & 12.48     & 15        & SE                     & QP                            & -                                       \\
\pplus{B1937}{21}    & 12.77     & 3         & QP + QP                & SE                            & -                                       \\
\pplus{J0030}{0451}  & 12.43     & 15        & QP                     & QP                            & -                                       \\
\pminus{J0613}{0200} & 12.25     & 15        & QP                     & -                             & AV, Dip ($\nu^{-4}$)                    \\
\pplus{J0645}{5158}  & 6.05      & 7         & SE                     & QP                            & -                                       \\
\pplus{J1012}{5307}  & 12.86     & 15        & QP                     & SE                            & -                                       \\
\pminus{J1024}{0719} & 7.68      & 15        & SE                     & -                             & -                                       \\
\pminus{J1455}{3330} & 12.88     & 7         & Ridge                  & -                             & -                                       \\
\pminus{J1600}{3053} & 9.64      & 15        & QP\_RF                 & QP                            & -                                       \\
\pminus{J1614}{2230} & 8.75      & 15        & QP                     & SE                            & AV                                      \\
\pplus{J1640}{2224}  & 12.35     & 15        & QP                     & SE                            & -                                       \\
\pplus{J1713}{0747}  & 12.43     & 3         & QP\_RF                 & QP                            & Dip ($\nu^{-2}$), Dip ($\nu^{-\chi}$)   \\
\pplus{J1738}{0333}  & 7.61      & 15        & SE                     & SE                            & -                                       \\
\pplus{J1741}{1351}  & 5.88      & 15        & SE                     & -                             & -                                       \\
\pminus{J1744}{1134} & 12.88     & 15        & SE                     & SE                            & -                                       \\
\pminus{J1909}{3744} & 12.69     & 15        & QP                     & QP                            & -                                        \\
\pplus{J1910}{1256}  & 8.30      & 15        & SE                     & -                             & -                                       \\
\pminus{J2010}{1323} & 7.76      & 15        & QP                     & -                             & -                                       \\
\pplus{J2043}{1711}  & 5.95      & 15        & QP                     & -                             & Cusp ($\nu^{-2}$), Cusp ($\nu^{-4}$)    \\
\pplus{J2317}{1439}  & 12.53     & 15        & QP                     & SE                            & Cusp ($\nu^{-2}$)                    \\
\hline
\end{tabular}
\caption{Preferred noise model for each pulsar.} \label{tab:modelresults} 
\end{table*}

%% file: acks12p5yrCNM.tex
The NANOGrav collaboration receives support from National Science Foundation (NSF) Physics Frontiers Center award \#2020265, the Gordon and Betty Moore Foundation, an NSERC Discovery Grant, and CIFAR.
The Arecibo Observatory is a facility of the NSF operated under cooperative agreement (AST-1744119) by the University of Central Florida (UCF) in alliance with Universidad Ana G. M\'endez (UAGM) and Yang Enterprises (YEI), Inc. The Green Bank Observatory is a facility of the NSF operated under cooperative agreement by Associated Universities, Inc. The National Radio Astronomy Observatory is a facility of the NSF operated under cooperative agreement by Associated Universities, Inc.
J.S.H. and J.B.\ acknowledge support from NSF CAREER award \#2339728.
P.R.B.\ is supported by the Science and Technology Facilities Council, grant number ST/W000946/1.
S.B.\ gratefully acknowledges the support of a Sloan Fellowship, and the support of NSF under award \#1815664.
H.T.C.\ acknowledges funding from the U.S. Naval Research Laboratory.
M.E.D.\ acknowledges support from the Naval Research Laboratory by NASA under contract S-15633Y.
T.D.\ and M.T.L.\ received support by an NSF Astronomy and Astrophysics Grant (AAG) award number 2009468 during this work.
E.C.F.\ is supported by NASA under award number 80GSFC24M0006.
D.C.G.\ is supported by NSF Astronomy and Astrophysics Grant (AAG) award \#2406919.
A.D.J.\ and M.V.\ acknowledge support from the Caltech and Jet Propulsion Laboratory President's and Director's Research and Development Fund.
A.D.J.\ acknowledges support from the Sloan Foundation.
N.L.\ was supported by the Vanderbilt Initiative in Data Intensive Astrophysics (VIDA) Fellowship.
Part of this research was carried out at the Jet Propulsion Laboratory, California Institute of Technology, under a contract with the National Aeronautics and Space Administration (80NM0018D0004).
D.R.L.\ and M.A.M.\ are supported by NSF \#1458952.
M.A.M.\ is supported by NSF \#2009425.
C.M.F.M.\ was supported in part by the National Science Foundation under Grants No.\ NSF PHY-1748958 and AST-2106552.
The Dunlap Institute is funded by an endowment established by the David Dunlap family and the University of Toronto.
T.T.P.\ acknowledges support from the Extragalactic Astrophysics Research Group at E\"{o}tv\"{o}s Lor\'{a}nd University, funded by the E\"{o}tv\"{o}s Lor\'{a}nd Research Network (ELKH), which was used during the development of this research.
S.M.R.\ and I.H.S.\ are CIFAR Fellows.
Portions of this work performed at NRL were supported by ONR 6.1 basic research funding.
The work of X.S.\ and J.P.S.\ is partly supported by the George and Hannah Bolinger Memorial Fund in the College of Science at Oregon State University.
J.S.\ is supported by an NSF Astronomy and Astrophysics Postdoctoral Fellowship under award AST-2202388, and acknowledges previous support by the NSF under award 1847938.
Pulsar research at UBC is supported by an NSERC Discovery Grant and by CIFAR.
S.R.T.\ acknowledges support from NSF AST-2007993.
S.R.T.\ acknowledges support from an NSF CAREER award \#2146016.
S.J.V.\ is supported by NSF award PHY-2011772.

%% file: Table_GP_Params_alt2.tex
\begin{table}
    \begin{center}
    \bgroup
    \setlength\tabcolsep{0.8mm}
    \begin{tabular}{ l | c c c c @{}}
        \hline Pulsar & $\log_{10}\sigma^{\rm{DM}}$ & $\log_{10}\ell^{\rm{DM}}$ & $\log_{10}\Gamma_p^{\rm{DM}}$ & $\log_{10}p^{\rm{DM}}$ \\
        \hline \pplus{B1855}{09} & $-6.25_{-0.10}^{+0.12}$ & $2.34_{-0.09}^{+0.09}$ & $ - $ & $ - $ \\
        \pplus{B1937}{21} & $-6.06_{-0.08}^{+0.06}$ & $1.96_{-0.04}^{+0.04}$ & $-1.86_{-0.10}^{+0.17}$ & $-1.7_{-0.2}^{+0.2}$ \\
        \pplus{B1937}{21} & $-4.19_{-0.06}^{+0.05}$ & $3.9_{-0.5}^{+0.7}$ & $-1.2_{-0.5}^{+1.1}$ & $1.1_{-0.2}^{+0.4}$ \\
        \pplus{J0030}{0451} & $-6.5_{-0.3}^{+0.4}$ & $3.0_{-0.2}^{+0.3}$ & $-1.2_{-1.1}^{+0.8}$ & $-1.1_{-0.6}^{+1.1}$ \\
        \pminus{J0613}{0200} & $-4.19_{-0.12}^{+0.07}$ & $3.8_{-0.2}^{+0.1}$ & $-1.6_{-0.4}^{+0.2}$ & $0.66_{-0.01}^{+0.02}$ \\
        \pplus{J0645}{5158} & $-5.9_{-0.6}^{+1.3}$ & $2.8_{-0.3}^{+0.3}$ & $ - $ & $ - $ \\
        \pplus{J1012}{5307} & $-6.77_{-0.07}^{+0.07}$ & $1.2_{-0.1}^{+0.5}$ & $0.3_{-2.2}^{+1.3}$ & $-0.5_{-1.0}^{+0.9}$ \\
        \pminus{J1024}{0719} & $-6.6_{-0.1}^{+0.2}$ & $2.3_{-0.4}^{+0.2}$ & $ - $ & $ - $ \\
        \pminus{J1455}{3330} & $-6.63_{-0.13}^{+0.09}$ & $ - $ & $ - $ & $ - $ \\
        \pminus{$^*$J1600}{3053} & $-6.27_{-0.05}^{+0.06}$ & $1.83_{-0.06}^{+0.06}$ & $-0.9_{-0.3}^{+0.2}$ & $-1.5_{-0.5}^{+0.3}$ \\
        \pminus{J1614}{2230} & $-6.06_{-0.05}^{+0.05}$ & $1.81_{-0.08}^{+0.06}$ & $0.2_{-0.1}^{+0.2}$ & $-1.2_{-0.6}^{+0.2}$ \\
        \pplus{J1640}{2224} & $-6.97_{-0.09}^{+0.10}$ & $1.89_{-0.19}^{+0.09}$ & $-0.2_{-0.3}^{+0.2}$ & $-1.2_{-0.6}^{+0.2}$ \\
        \pplus{$^*$J1713}{0747} & $-7.06_{-0.04}^{+0.05}$ & $1.58_{-0.09}^{+0.22}$ & $-0.2_{-0.4}^{+0.2}$ & $-1.08_{-0.03}^{+0.02}$ \\
        \pplus{J1738}{0333} & $-6.3_{-0.3}^{+0.4}$ & $2.4_{-0.5}^{+0.2}$ & $ - $ & $ - $ \\
        \pplus{J1741}{1351} & $-6.8_{-0.1}^{+0.1}$ & $1.99_{-0.08}^{+0.09}$ & $ - $ & $ - $ \\
        \pminus{J1744}{1134} & $-6.4_{-0.3}^{+0.5}$ & $2.9_{-0.2}^{+0.2}$ & $ - $ & $ - $ \\
        \pminus{J1909}{3744} & $-6.97_{-0.05}^{+0.04}$ & $1.51_{-0.10}^{+0.13}$ & $0.0_{-0.3}^{+0.3}$ & $-1.1_{-0.6}^{+0.1}$\\
        \pplus{J1910}{1256} & $-6.0_{-0.2}^{+0.3}$ & $2.6_{-0.1}^{+0.2}$ & $ - $ & $ - $ \\
        \pminus{J2010}{1323} & $-6.45_{-0.08}^{+0.10}$ & $2.05_{-0.08}^{+0.07}$ & $-0.9_{-0.4}^{+0.3}$ & $-0.8_{-0.9}^{+0.1}$ \\
        \pplus{J2043}{1711} & $-7.1_{-0.1}^{+0.1}$ & $2.0_{-0.1}^{+0.5}$ & $-0.8_{-1.2}^{+1.4}$ & $-0.1_{-1.3}^{+0.8}$ \\
        \pplus{J2317}{1439} & $-6.34_{-0.07}^{+0.08}$ & $2.17_{-0.05}^{+0.07}$ & $-1.5_{-0.3}^{+0.4}$ & $-0.4_{-1.1}^{+0.2}$ \\
        \hline Pulsar & $\log_{10}\sigma^{\rm{Chr}}$ & $\log_{10}\ell^{\rm{Chr}}$ & $\log_{10}\Gamma_p^{\rm{Chr}}$ & $\log_{10}p^{\rm{Chr}}$ \\
        \hline \pplus{B1855}{09} & $-7.84_{-0.08}^{+0.08}$ & $1.7_{-0.3}^{+0.2}$ & $0.4_{-0.4}^{+1.0}$ & $-1.1_{-0.6}^{+0.1}$ \\
        \pplus{B1937}{21} & $-4.38_{-0.04}^{+0.03}$ & $4.4_{-0.2}^{+0.1}$ & $ - $ & $ - $ \\
        \pplus{J0030}{0451} & $-7.79_{-0.05}^{+0.06}$ & $1.6_{-0.4}^{+1.1}$ & $1.7_{-0.6}^{+0.2}$ & $-1.2_{-0.4}^{+0.5}$ \\
        \pplus{J0645}{5158} & $-7.6_{-0.1}^{+0.1}$ & $1.7_{-0.4}^{+1.4}$ & $0.1_{-2.0}^{+1.4}$ & $-0.4_{-1.2}^{+0.9}$ \\
        \pplus{J1012}{5307} & $-7.1_{-0.1}^{+0.1}$ & $2.1_{-0.2}^{+0.4}$ & $ - $ & $ - $ \\
        \pminus{J1600}{3053} & $-6.82_{-0.07}^{+0.07}$ & $1.9_{-0.2}^{+0.3}$ & $-0.1_{-0.3}^{+0.3}$ & $-1.2_{-0.6}^{+0.3}$ \\
        \pminus{J1614}{2230} & $-7.7_{-1.5}^{+0.6}$ & $1.9_{-0.7}^{+1.5}$ & $ - $ & $ - $ \\
        \pplus{J1640}{2224} & $-7.6_{-0.2}^{+0.2}$ & $2.7_{-0.1}^{+0.1}$ & $ - $ & $ - $ \\
        \pplus{J1713}{0747} & $-7.45_{-0.06}^{+0.05}$ & $1.7_{-0.2}^{+0.2}$ & $1.3_{-0.3}^{+0.3}$ & $-1.10_{-0.01}^{+0.01}$ \\
        \pplus{J1738}{0333} & $-5.6_{-0.5}^{+0.5}$ & $3.5_{-0.7}^{+0.4}$ & $ - $ & $ - $ \\
        \pminus{J1744}{1134} & $-7.36_{-0.06}^{+0.06}$ & $1.2_{-0.1}^{+0.1}$ & $ - $ & $ - $ \\
        \pminus{J1909}{3744} & $-7.71_{-0.07}^{+0.07}$ & $1.4_{-0.3}^{+0.7}$ & $1.2_{-1.9}^{+0.6}$ & $-0.7_{-1.0}^{+0.6}$ \\
        \pplus{J2317}{1439} & $-5.74_{-0.07}^{+0.08}$ & $3.93_{-0.10}^{+0.05}$ & $ - $ & $ - $ \\
        \hline
    \end{tabular}
    \egroup
    \caption{DM and chromatic GP noise parameter medians and 68.3\% Bayesian credible intervals. Dashes are used in place of parameters that were not included in the preferred model. Here, \psrplus{B1937}{21} is listed twice since its preferred model includes two DM GPs. Not all pulsars include an additional chromatic model. $^*$These two pulsars favor the QP\_RF kernel and each includes 2 additional DMGP parameters. The additional parameters are for \psrminus{J1600}{3053}, $\log_{10}\alpha_{\mathrm{wgt}} = -1.0_{-0.5}^{+1.0}$, and $\log_{10}\ell_2 = 2.6_{-0.4}^{+0.4}$; and for \psrplus{J1713}{0747}, $\log_{10}\alpha_{\mathrm{wgt}} = -1.4_{-0.4}^{+0.6}$, and $\log_{10}\ell_2 = 2.6_{-0.5}^{+0.4}$.}
    \label{tab:chrom_params}
    \end{center}
\end{table}

%% file: Table_DetParams.tex
\begin{table*}
    \centering
    \begin{tabular}{ l | c c c c c | c c @{}}
        \hline Pulsar & $\log_{10}A^{\rm{dip/cusp}}$ & $\log_{10}\tau_{\rm{pre}}^{\rm{dip/cusp}}$ & $\log_{10}\tau_{\rm{post}}^{\rm{dip/cusp}}$ & $t_0^{\rm{dip/cusp}}$ & $\chi^{\rm{dip/cusp}}$ & $\log_{10}\sigma^{\rm{AV}}$ & $\log_{10}\phi^{\rm{AV}}$ \\
        \hline \pminus{J0613}{0200} & $-6.9_{-0.2}^{+0.4}$ & $ - $ & $0.8_{-0.5}^{+1.0}$ & $56487.4_{-1.2}^{+1432.6}$ & 4 & $-6.73_{-0.07}^{+0.06}$ & $2.4_{-0.1}^{+0.1}$ \\
        \pminus{J1614}{2230} & $ - $ & $ - $ & $ - $ & $ - $ & $ - $ & $-7.8_{-1.5}^{+1.2}$ & $3.0_{-1.5}^{+1.6}$ \\
        \pplus{J1713}{0747} & $-5.82_{-0.04}^{+0.04}$ & $ - $ & $2.00_{-0.08}^{+0.08}$ & $54758.6_{-4.6}^{+4.8}$ & 2 & $ - $ & $ - $ \\
        \pplus{J1713}{0747} & $-5.85_{-0.03}^{+0.03}$ & $ - $ & $1.56_{-0.07}^{+0.07}$ & $57510.3_{-1.3}^{+1.3}$ & $1.35_{-0.08}^{+0.08}$ & $ - $ & $ - $ \\
        \pplus{J2043}{1711} & $-7.1_{-0.4}^{+0.6}$ & $0.04_{-0.03}^{+0.07}$ & $0.2_{-0.2}^{+0.3}$ & $57377.0_{-1.0}^{+1.7}$ & 2 & $ - $ & $ - $ \\
        \pplus{J2043}{1711} & $-6.67_{-0.12}^{+0.10}$ & $0.7_{-0.4}^{+0.4}$ & $1.8_{-0.4}^{+0.5}$ & $57084.7_{-3.9}^{+2.3}$ & 4 & $ - $ & $ - $ \\
        \pplus{J2317}{1439} & $-6.64_{-0.09}^{+0.26}$ & $2.0_{-0.5}^{+0.4}$ & $0.2_{-0.1}^{+0.4}$ & $53385.7_{-0.3}^{+3699.2}$ & 2 & $ - $ & $ - $ \\
        \hline
    \end{tabular}
    \caption{Deterministic parameter medians and 68.3\% Bayesian credible intervals. Includes exponential dip/cusp events and Annual DM Variations (AV). Exponential dips, which are special cases of the generic cusp model, do not include $\log_{10}\tau_{\rm{pre}}$. Here the preferred models for \psrplus{J1713}{0747} and \psrplus{J2043}{1711} favored multiple events.}
    \label{tab:det_params}
\end{table*}

%% file: Table_RN_Params.tex
\begin{table}
    \centering
    \setlength{\tabcolsep}{0.4em}
    \begin{tabular}{ l | c c | c c @{}}
        & \multicolumn{2}{c}{Standard noise model} & \multicolumn{2}{c}{Custom noise model} \\
        \hline Pulsar & $\log_{10}A_{\rm{RN}}$ & $\gamma_{\rm{RN}}$ & $\log_{10}A_{\rm{RN}}$ & $\gamma_{\rm{RN}}$ \\
        \hline \pplus{B1855}{09} & $-14.0_{-0.4}^{+0.4}$ & $4.2_{-1.1}^{+1.2}$ & $-14.0_{-0.5}^{+0.4}$ & $4.2_{-1.0}^{+1.2}$ \\
        \pplus{B1937}{21} & $-13.5_{-0.1}^{+0.1}$ & $3.4_{-0.4}^{+0.5}$ & $-13.4_{-0.1}^{+0.1}$ & $3.2_{-0.3}^{+0.4}$ \\
        \pplus{J0030}{0451} & $-14.7_{-0.5}^{+0.5}$ & $5.3_{-1.2}^{+1.1}$ & $-14.5_{-0.5}^{+0.5}$ & $4.9_{-1.1}^{+1.2}$ \\
        \pminus{J0613}{0200} & $-13.5_{-0.6}^{+0.2}$ & $2.1_{-0.8}^{+1.4}$ & $-13.7_{-0.9}^{+0.3}$ & $2.5_{-0.9}^{+1.9}$ \\
        \pplus{J0645}{5158} & $-14.2^{95\%}$ & $ - $ & $-13.7^{95\%}$ & $ - $ \\
        \pplus{J1012}{5307} & $-12.8_{-0.1}^{+0.1}$ & $1.3_{-0.4}^{+0.4}$ & $-13.1_{-0.1}^{+0.1}$ & $1.3_{-0.4}^{+0.4}$ \\
        \pminus{J1024}{0719} & $-13.0^{95\%}$ & $ - $ & $-13.5^{95\%}$ & $ - $ \\
        \pminus{J1455}{3330} & $-13.7^{95\%}$ & $ - $ & $-13.4_{-0.8}^{+0.3}$ & $2.2_{-1.0}^{+2.0}$ \\
        \pminus{J1600}{3053} & $-13.3_{-0.1}^{+0.1}$ & $0.3_{-0.2}^{+0.5}$ & $-13.3^{95\%}$ & $ - $ \\
        \pminus{J1614}{2230} & $-14.4^{95\%}$ & $ - $ & $-14.4^{95\%}$ & $ - $ \\
        \pplus{J1640}{2224} & $-14.1^{95\%}$ & $ - $ & $-14.2^{95\%}$ & $ - $ \\
        \pplus{J1713}{0747} & $-14.0_{-0.1}^{+0.1}$ & $1.1_{-0.5}^{+0.5}$ & $-15.3_{-0.9}^{+0.7}$ & $4.2_{-1.5}^{+1.8}$ \\
        \pplus{J1738}{0333} & $-14.2^{95\%}$ & $ - $ & $-13.3^{95\%}$ & $ - $ \\
        \pplus{J1741}{1351} & $-13.6^{95\%}$ & $ - $ & $-13.7^{95\%}$ & $ - $ \\
        \pminus{J1744}{1134} & $-13.6_{-1.1}^{+0.3}$ & $2.9_{-1.0}^{+2.2}$ & $-14.9_{-0.5}^{+0.6}$ & $5.5_{-1.3}^{+1.0}$ \\
        \pminus{J1909}{3744} & $-14.8_{-0.7}^{+0.7}$ & $4.4_{-1.6}^{+1.6}$ & $-14.3_{-0.5}^{+0.3}$ & $3.7_{-0.8}^{+1.3}$ \\
        \pplus{J1910}{1256} & $-13.7^{95\%}$ & $ - $ & $-13.2^{95\%}$ & $ - $ \\
        \pminus{J2010}{1323} & $-14.1^{95\%}$ & $ - $ & $-13.8^{95\%}$ & $ - $ \\
        \pplus{J2043}{1711} & $-14.2_{-0.7}^{+0.4}$ & $3.7_{-1.5}^{+1.9}$ & $-13.9^{95\%}$ & $ - $ \\
        \pplus{J2317}{1439} & $-15.1_{-0.7}^{+0.8}$ & $5.3_{-1.6}^{+1.2}$ & $-15.0_{-0.6}^{+0.7}$ & $5.4_{-1.6}^{+1.1}$ \\
        \hline
    \end{tabular}
    \caption{Red noise parameter medians and 68.3\% Bayesian credible intervals, under both the standard and customized noise models. Where the red noise process is statistically insignificant (BF$^{\rm{RN}}$ $< 10$) under the model, we instead report the 95\% upper limit (one-sided Bayesian credible interval) on the red noise amplitude.}
    \label{tab:rn_params}
\end{table}

%% file: Table_ECORR_Params.tex
\begin{table*}
    \centering
    \begin{minipage}{.49\linewidth}
        \centering
        \bgroup
        \setlength\tabcolsep{1.0mm}
        \begin{tabular}{ l | l | c | c @{}}
            \hline Pulsar & Rcvr/Backend & Std. noise & Cus. noise \\
            \hline \multirow{4}{*}{\pplus{B1855}{09}} & 430\_ASP & $-6.4^{95\%}$ & $-6.4^{95\%}$ \\
             & 430\_PUPPI & $-5.5^{95\%}$ & $-6.1^{95\%}$ \\
             & L-wide\_ASP & $-6.09_{-0.05}^{+0.05}$ & $-6.09_{-0.05}^{+0.05}$ \\
             & L-wide\_PUPPI & $-6.64_{-0.07}^{+0.07}$ & $-6.57_{-0.04}^{+0.04}$ \\
            \hline \multirow{8}{*}{\pplus{B1937}{21}} & \textbf{L-wide\_ASP} & $\mathbf{-6.94_{-0.10}^{+0.09}}$ & $\mathbf{-6.9^{95\%}}$ \\
             & L-wide\_PUPPI & $-7.04_{-0.07}^{+0.07}$ & $-7.28_{-0.13}^{+0.10}$ \\
             & \textbf{Rcvr1\_2\_GASP} & $\mathbf{-6.96_{-0.06}^{+0.05}}$ & $\mathbf{-7.4^{95\%}}$ \\
             & Rcvr1\_2\_GUPPI & $-6.87_{-0.05}^{+0.05}$ & $-7.04_{-0.05}^{+0.05}$ \\
             & Rcvr\_800\_GASP & $-7.4^{95\%}$ & $-7.5^{95\%}$ \\
             & Rcvr\_800\_GUPPI & $-6.44_{-0.05}^{+0.04}$ & $-6.47_{-0.05}^{+0.05}$ \\
             & S-wide\_ASP & $-6.56_{-0.07}^{+0.07}$ & $-6.70_{-0.09}^{+0.09}$ \\
             & S-wide\_PUPPI & $-6.87_{-0.09}^{+0.09}$ & $-6.91_{-0.09}^{+0.09}$ \\
            \hline \multirow{5}{*}{\pplus{J0030}{0451}} & 430\_ASP & $-6.7^{95\%}$ & $-6.8^{95\%}$ \\
             & \textbf{430\_PUPPI} & $\mathbf{-6.3^{95\%}}$ & $\mathbf{-6.5_{-0.1}^{+0.1}}$ \\
             & L-wide\_ASP & $-6.8^{95\%}$ & $-6.8^{95\%}$ \\
             & L-wide\_PUPPI & $-7.0^{95\%}$ & $-6.8^{95\%}$ \\
             & S-wide\_PUPPI & $-6.5^{95\%}$ & $-6.5^{95\%}$ \\
            \hline \multirow{4}{*}{\pminus{J0613}{0200}} & Rcvr1\_2\_GASP & $-6.6^{95\%}$ & $-6.5^{95\%}$ \\
             & Rcvr1\_2\_GUPPI & $-6.6^{95\%}$ & $-6.6^{95\%}$ \\
             & Rcvr\_800\_GASP & $-6.9^{95\%}$ & $-6.8^{95\%}$ \\
             & \textbf{Rcvr\_800\_GUPPI} & $\mathbf{-6.8_{-0.1}^{+0.1}}$ & $\mathbf{-6.7^{95\%}}$ \\
            \hline \multirow{2}{*}{\pplus{J0645}{5158}} & Rcvr1\_2\_GUPPI & $-6.9^{95\%}$ & $-6.7^{95\%}$ \\
             & Rcvr\_800\_GUPPI & $-6.8^{95\%}$ & $-6.9^{95\%}$ \\
            \hline \multirow{4}{*}{\pplus{J1012}{5307}} & Rcvr1\_2\_GASP & $-6.2^{95\%}$ & $-6.2^{95\%}$ \\
             & Rcvr1\_2\_GUPPI & $-6.39_{-0.06}^{+0.06}$ & $-6.36_{-0.05}^{+0.05}$ \\
             & Rcvr\_800\_GASP & $-6.8^{95\%}$ & $-6.8^{95\%}$ \\
             & Rcvr\_800\_GUPPI & $-6.9^{95\%}$ & $-6.8^{95\%}$ \\
            \hline \multirow{4}{*}{\pminus{J1024}{0719}} & Rcvr1\_2\_GASP & $-5.8^{95\%}$ & $-5.8^{95\%}$ \\
             & Rcvr1\_2\_GUPPI & $-6.4^{95\%}$ & $-6.5^{95\%}$ \\
             & Rcvr\_800\_GASP & $-5.6^{95\%}$ & $-5.8^{95\%}$ \\
             & \textbf{Rcvr\_800\_GUPPI} & $\mathbf{-6.5^{95\%}}$ & $\mathbf{-6.53_{-0.13}^{+0.10}}$ \\
            \hline \multirow{4}{*}{\pminus{J1455}{3330}} & Rcvr1\_2\_GASP & $-5.6^{95\%}$ & $-5.6^{95\%}$ \\
             & Rcvr1\_2\_GUPPI & $-6.7^{95\%}$ & $-6.6^{95\%}$ \\
             & Rcvr\_800\_GASP & $-5.9^{95\%}$ & $-6.0^{95\%}$ \\
             & Rcvr\_800\_GUPPI & $-5.8^{95\%}$ & $-6.6^{95\%}$ \\
            \hline \multirow{4}{*}{\pminus{J1600}{3053}} & Rcvr1\_2\_GASP & $-6.4^{95\%}$ & $-6.7^{95\%}$ \\
             & \textbf{Rcvr1\_2\_GUPPI} & $\mathbf{-6.9_{-0.2}^{+0.1}}$ & $\mathbf{-6.9^{95\%}}$ \\
             & Rcvr\_800\_GASP & $-6.3^{95\%}$ & $-6.5^{95\%}$ \\
             & \textbf{Rcvr\_800\_GUPPI} & $\mathbf{-6.29_{-0.11}^{+0.08}}$ & $\mathbf{-6.7^{95\%}}$ \\
            \hline \multirow{4}{*}{\pminus{J1614}{2230}} & Rcvr1\_2\_GASP & $-6.4^{95\%}$ & $-6.4^{95\%}$ \\
             & Rcvr1\_2\_GUPPI & $-7.0^{95\%}$ & $-7.0^{95\%}$ \\
             & Rcvr\_800\_GASP & $-6.2^{95\%}$ & $-6.2^{95\%}$ \\
             & Rcvr\_800\_GUPPI & $-6.5^{95\%}$ & $-6.2^{95\%}$ \\
             \hline
        \end{tabular}
        \egroup
    \end{minipage}
    \begin{minipage}{.49\linewidth}
        \centering
        \bgroup
        \setlength\tabcolsep{1.0mm}
        \begin{tabular}{ l | l | c | c @{}}
            \hline Pulsar & Rcvr/Backend & Std. noise & Cus. noise \\
            \hline \multirow{4}{*}{\pplus{J1640}{2224}} & 430\_ASP & $-7.0^{95\%}$ & $-7.0^{95\%}$ \\
             & \textbf{430\_PUPPI} & $\mathbf{-6.3^{95\%}}$ & $\mathbf{-6.25_{-0.04}^{+0.04}}$ \\
             & L-wide\_ASP & $-6.28_{-0.08}^{+0.08}$ & $-6.32_{-0.08}^{+0.08}$ \\
             & L-wide\_PUPPI & $-6.41_{-0.03}^{+0.03}$ & $-6.38_{-0.03}^{+0.03}$ \\
            \hline \multirow{8}{*}{\pplus{J1713}{0747}} & L-wide\_ASP & $-7.03_{-0.08}^{+0.08}$ & $-6.94_{-0.07}^{+0.07}$ \\
             & L-wide\_PUPPI & $-7.14_{-0.04}^{+0.04}$ & $-7.13_{-0.03}^{+0.03}$ \\
             & Rcvr1\_2\_GASP & $-7.0^{95\%}$ & $-7.1^{95\%}$ \\
             & Rcvr1\_2\_GUPPI & $-7.13_{-0.03}^{+0.03}$ & $-7.19_{-0.03}^{+0.03}$ \\
             & Rcvr\_800\_GASP & $-7.1^{95\%}$ & $-7.1^{95\%}$ \\
             & Rcvr\_800\_GUPPI & $-6.73_{-0.07}^{+0.06}$ & $-6.94_{-0.08}^{+0.08}$ \\
             & S-wide\_ASP & $-6.88_{-0.08}^{+0.08}$ & $-6.86_{-0.08}^{+0.08}$ \\
             & S-wide\_PUPPI & $-7.30_{-0.04}^{+0.04}$ & $-7.27_{-0.04}^{+0.04}$ \\
            \hline \multirow{4}{*}{\pplus{J1738}{0333}} & L-wide\_ASP & $-6.1^{95\%}$ & $-6.4^{95\%}$ \\
             & \textbf{L-wide\_PUPPI} & $ \mathbf{-6.8^{95\%}}$ & $\mathbf{-6.85_{-0.16}^{+0.10}}$ \\
             & S-wide\_ASP & $-6.3^{95\%}$ & $-6.4^{95\%}$ \\
             & S-wide\_PUPPI & $-6.9^{95\%}$ & $-6.9^{95\%}$ \\
            \hline \multirow{4}{*}{\pplus{J1741}{1351}} & 430\_ASP & $-5.1^{95\%}$ & $-5.2^{95\%}$ \\
             & \textbf{430\_PUPPI} & $\mathbf{-6.1^{95\%}}$ & $\mathbf{-6.24_{-0.07}^{+0.07}}$ \\
             & L-wide\_ASP & $-6.1^{95\%}$ & $-6.1^{95\%}$ \\
             & L-wide\_PUPPI & $-6.9_{-0.1}^{+0.1}$ & $-6.92_{-0.10}^{+0.08}$ \\
            \hline \multirow{4}{*}{\pminus{J1744}{1134}} & Rcvr1\_2\_GASP & $-6.29_{-0.10}^{+0.10}$ & $-6.37_{-0.09}^{+0.09}$ \\
             & Rcvr1\_2\_GUPPI & $-6.54_{-0.05}^{+0.05}$ & $-6.57_{-0.03}^{+0.04}$ \\
             & \textbf{Rcvr\_800\_GASP} & $\mathbf{-6.4^{95\%}}$ & $\mathbf{-6.41_{-0.11}^{+0.09}}$ \\
             & Rcvr\_800\_GUPPI & $-6.40_{-0.06}^{+0.06}$ & $-6.42_{-0.05}^{+0.05}$ \\
            \hline \multirow{4}{*}{\pminus{J1909}{3744}} & Rcvr1\_2\_GASP & $-7.5^{95\%}$ & $-7.5^{95\%}$ \\
             & Rcvr1\_2\_GUPPI & $-7.14_{-0.03}^{+0.03}$ & $-7.10_{-0.03}^{+0.03}$ \\
             & Rcvr\_800\_GASP & $-7.3^{95\%}$ & $-7.2^{95\%}$ \\
             & Rcvr\_800\_GUPPI & $-7.27_{-0.08}^{+0.08}$ & $-6.98_{-0.07}^{+0.06}$ \\
            \hline \multirow{4}{*}{\pplus{J1910}{1256}} & L-wide\_ASP & $-6.1^{95\%}$ & $-6.2^{95\%}$ \\
             & \textbf{L-wide\_PUPPI} & $\mathbf{-6.5^{95\%}}$ & $\mathbf{-6.70_{-0.12}^{+0.09}}$ \\
             & S-wide\_ASP & $-6.2^{95\%}$ & $-6.2^{95\%}$ \\
             & S-wide\_PUPPI & $-6.22_{-0.08}^{+0.07}$ & $-6.26_{-0.08}^{+0.07}$ \\
            \hline \multirow{4}{*}{\pminus{J2010}{1323}} & Rcvr1\_2\_GASP & $-5.9^{95\%}$ & $-5.9^{95\%}$ \\
             & Rcvr1\_2\_GUPPI & $-6.9^{95\%}$ & $-7.0^{95\%}$ \\
             & Rcvr\_800\_GASP & $-5.9^{95\%}$ & $-6.2^{95\%}$ \\
             & Rcvr\_800\_GUPPI & $-6.8^{95\%}$ & $-6.5^{95\%}$ \\
            \hline \multirow{4}{*}{\pplus{J2043}{1711$^*$}} & 430\_ASP & $-5.9^{95\%}$ & $-6.2^{95\%}$ \\
             & \textbf{430\_PUPPI} & $\mathbf{-6.7^{95\%}}$ & $\mathbf{-6.35_{-0.04}^{+0.04}}$ \\
             & L-wide\_ASP & $-6.2^{95\%}$ & $-6.1^{95\%}$ \\
             & \textbf{L-wide\_PUPPI} & $\mathbf{-7.1^{95\%}}$ & $\mathbf{-7.01_{-0.09}^{+0.08}}$ \\
            \hline \multirow{5}{*}{\pplus{J2317}{1439}} & 327\_ASP & $-6.9^{95\%}$ & $-6.4^{95\%}$ \\
             & 327\_PUPPI & $-6.1^{95\%}$ & $-6.5^{95\%}$ \\
             & 430\_ASP & $-6.51_{-0.07}^{+0.06}$ & $-6.5_{-0.1}^{+0.1}$ \\
             & \textbf{430\_PUPPI} & $\mathbf{-6.4^{95\%}}$ & $\mathbf{-6.25_{-0.04}^{+0.04}}$ \\
             & L-wide\_PUPPI & $-6.65_{-0.05}^{+0.05}$ & $-6.66_{-0.05}^{+0.05}$ \\
             \hline
        \end{tabular}
        \egroup
    \end{minipage}
    \caption{$\log_{10}\rm{ECORR}$ medians and 68.3\% Bayesian credible intervals, under both the standard and customized noise models. ECORR params are reported by pulsar and backend. Where ECORR is statistically insignificant (BF$^{\rm{ECORR}}$ $< 10$) under the model, we instead report the 95\% upper limit (one-sided Bayesian credible interval) on $\log_{10}\rm{ECORR}$. We bold all backends/ECORR parameters that change from significant to insignificant or vice versa, corresponding to Figure~\ref{fig:ecorr_posteriors}. 
    $^*$See Figure~\ref{fig:j2043_ecorr} for a discussion of how additional changes to the model make the ECORRS for \psrplus{J2043}{1711} insignificant.}
    \label{tab:ecorr_params}
\end{table*}